KYIV NATIONAL TARAS SHEVCHENKO UNIVERSITY

On the rights of the manuscript

BOGDAN M. TANYGIN

# MAGNETOELECTRIC PROPERTIES AND SYMMETRY CLASSIFICATION OF MICROMAGNETIC STRUCTURES

01.04.11 - magnetism

Ph.D. Thesis
physical and mathematical sciences

Scientific supervisor
Kovalenko Valery Fadeevich
Doctor of Physical and Mathematical Sciences, Professor

Kyiv — 2012

*Machine translation from Ukrainian by DeepL AI;*
*revised by Dr. Bogdan Tanygin (2023)*

# TABLE OF CONTENT

# LIST OF ABBREVIATIONS

DW - domain wall;

ME - magnetoelectric;

BL - Bloch line;

MV - magnetic vortex;

MAV - magnetic antivortex;

BP - Bloch point.

# INTRODUCTION

The development of magnetic memory technologies, spintronic and other electronics devices is significantly related [1,2] to the use of homogeneous [3-6] and inhomogeneous [7-9] magnetoelectric (ME) effects, i.e. the influence of an external electrostatic field on the magnetization of a material and/or a similar effect of an external magnetic field on electric polarization. These effects are caused by ME interactions of the electrical and magnetic subsystems in a solid. Among the applications of ME effects, the methods of magnetic reversal under the influence of an electrostatic field and the reverse effect — magnetic capacity (the effect of a magnetic field on the dielectric constant) are of primary interest. This makes it possible to change the state of memory elements, spin valves, logic elements, including the possibility of developing 4-state logic, in computing, to design new ME types of sensors, etc. Nonvolatile memory such as MRAM (magnetoresistive random access memory) is promising [10]. The advantage of direct electrostatic influence (on magnetic structures) over conventional methods of spintronics [11] is that there is no need to use electric currents to create a magnetic field or transfer spin polarization. The technological saturation of the development of existing methods is due to the fact that the reduction in the size of the elements ultimately leads to the maximum permissible current densities (of the order of and above $\sim 10^8$ $A/cm^2$), which cause technologically unacceptably large ohmic losses (unacceptably high heating) and gradual degradation of devices at room temperatures. For the ME control channel in electronics and spintronics, however, there is no limitation on the current density due to the absence of the current itself. This qualitatively expands the existing technological limits of element miniaturization (and information recording density) and process frequency increase, which plays an important role in the further development of the main elements of modern electronic computing.

**Relevance of the topic:**

In recent years (in fact, since the beginning of the 21st century), interest in magnetoelectrics has increased significantly [1,2,12-14]. Existing patents [15] for methods of recording information using ME materials have limitations on operating temperatures (requiring low temperatures). At room temperatures, ME control is possible in composite multiferroics [16], which is usually implemented by means of the coupling of the magnetic and electrical subsystems through magnetostriction and piezoelectric effect. However, the use of composite multiferroics has its drawbacks, including miniaturization problems, since the coexistence of different phases complicates the process of creating elements and creates additional problems of degradation of electronics elements.

In general, this area is in the process of active development, and a number of opportunities have not yet been utilized in technology. Thus, the ME control of micromagnetic structures [17-19] deserves attention. Among the latter, it is important to consider magnetic vortices (MVs) and antivortices (AVs) in nanomagnets [18,20], as well as localized magnetic solitons (skyrmions) [21]. Even without use of ME effects, there are patents and invention notices [22-26] on a magnetic memory based on such micromagnetic structures as, respectively, magnetic domain walls (MDWs) [22] (0° MDWs [27]), Bloch lines (BLs) [23], MVs and MAVs [24] (VRAM and AVRAM type memory, respectively), skyrmions [25], and Bloch points (BPs) [26]. It should be emphasized that skyrmions are of particular importance due to the possibility of their *collapse to atomic size* [28]. MVs are stabilized in crystals with chiral interactions or low-dimensional structures (magnetic dots). In the latter case, they play a role not only as memory elements but also as microwave signal generators.

The ME control mechanism of micromagnetic structures plays an important role in this type of applications due to the advantages of electrostatic control described above. Nevertheless, similar ME devices have not yet been developed, which confirms the insufficiency of experimental and theoretical studies of ME effects in micromagnetic structures.

The ME properties of magnetic DWs have been recorded only on the basis of indirect experimental evidence [29,30]. The direct experimental observation of the Néel wall displacement in the film $(BiLu)_3(FeGa)_5O_{12}$ under the influence of a strong gradient of the tip electrostatic field was observed only in 2008 [31]. An experimental result on the control of the BL by an electrostatic field was reported [32]. These experimental results in sufficiently studied materials and micromagnetic structures (domain walls) look like a "classical" result of magnetism physics, although chronologically the results are modern. Such a contradiction requires intensification of theoretical research in the relevant direction, and it is assumed that the existing theoretical framework needs to be supplemented.

For the first time in 1983, V.G. Baryakhtar et al. theoretically showed that magnetic DWs and other micromagnetic structures can have ME properties that manifest themselves as a *inhomogeneous ME effect* [7,9], i.e. the ME interaction of spatially inhomogeneous magnetization with the electric polarization of the material. The name is opposed to the usual *homogeneous ME effect,* i.e. the interaction of homogeneous magnetization and electric polarization [4-6]. Further theoretical studies have been carried out in various special cases of DW, MV, and other micromagnetic structures [33-36]. At the same time, group-theoretical methods play a key role in the study of ME effects. The symmetry classification of homogeneous ME effects was developed [6]. It not only indicates the symmetry of crystals in which a homogeneous ME effect is possible, but also allows predicting its anisotropy and describing it qualitatively.

The symmetric ME classification has been constructed for most $180^0$ DWs [37] and for BLs in them [38] without specifying (in the classification of BLs) the distribution of electric polarization for each magnetic group of symmetry. It makes it possible to establish a qualitative link between the spatial distribution of magnetization and the induced electric polarization in the volume of the DW without solving the variational problem, which requires numerical modeling in the most cases. At the same time, a purely qualitative symmetry analysis is sufficient to predict and analyze ME phenomena. The completeness of such a classification allows us to describe the properties of a whole group of physical objects, which, traditionally, in micromagnetism, are obtained only after the quantitative solution of each specific variational problem.

A similar symmetry ME classification for other DWs (zero-degree DWs and DWs of intermediate types) and other micromagnetic structures (BLs in the volume of any DW, BPs in the volume of any BLs, skyrmions, MVs, MAVs) has not yet been developed. At the same time, the need to study the ME properties of these micromagnetic structures is important for the development of modern electronics and spintronics devices. The need for such a classification exists, firstly, for reasons of the development of the fundamental theory of micromagnetism, since it is a symmetry classification of most micromagnetic structures. Secondly, such a classification makes it possible to predict and describe qualitatively inhomogeneous ME effects that can be observed experimentally in micromagnetic structures. This can be regarded as one of the effective applications of the group-theoretical approach.

The need for theoretical studies exists not only with respect to the symmetry of micromagnetic structures. Inhomogeneous ME interaction is possible in a magnetically ordered crystal of any symmetry [35]. At the same time, the phenomenological theory of inhomogeneous ME interaction for a crystal of a particular crystallographic symmetry was constructed for a special case of the variational problem for a cubic $m\bar{3}m$ crystal [33]. The phenomenological theory of inhomogeneous ME interaction is not extended to some point groups of the cubic crystal system ($432$, $\bar{4}3m$, $m\bar{3}$ and $23$) and other crystal systems. At the same time, it is obviously necessary for the analysis of experimental studies of inhomogeneous ME effects in micromagnetic structures. Without such an analysis, it is impossible to test the reliability of hypotheses regarding the microscopic mechanism of ME interaction (the relevant bibliography review is presented in this thesis) or plan experimental research of ME interactions in magnetic and multiferroic media.

**Linking the work to research programs, plans, and topics:**

The work was performed at the Department of Electrophysics, Faculty of Radiophysics, Taras Shevchenko National University of Kyiv.

The research was conducted and funded within the framework of state budget research topics:

1. "Investigation of the effects of interaction of electromagnetic and acoustic fields with ordered, nanostructured and biological systems for the creation of new technologies" (registration number 011U005265).
2. "Physical bases of the elemental base and effects of interaction of radiation with matter for the development of the latest information technologies" (registration number 0106U006545).

**The purpose and objectives of the research:**

*The aim of the* research carried out in this thesis is to develop a symmetry classification of the main micromagnetic structures (magnetic domain walls of any type, Bloch lines, Bloch points, and magnetic vortices in the sample plane), establishing its connection with the phenomenological theory of inhomogeneous ME interaction and applying (adapting) the phenomenological theory to magnetically ordered crystals of all symmetry groups, including cubic, tetragonal and orthorhombic crystal systems.

The goal was achieved by completing the tasks:

– to build (adapt) a phenomenological theory and systematize the corresponding microscopic mechanisms of inhomogeneous ME interactions in crystals belonging to all crystallographic groups of cubic, tetragonal and orthorhombic crystal systems;
– extend the symmetry DW classification to any type of magnetic DW (including 0° DW);

- extend the symmetry classification of BLs to BLs in the volume of arbitrary magnetic DWs;
- to build a symmetry classification of arbitrary BPs;
- to suggest a symmetry classification of the vortex magnetization distribution in the plates/films;
- to explain the essence of contradictions and imperfections in the definitions of chirality provided by different authors in the modern bibliography. Find ways to improve them;
- to establish a connection between the constructed general classifications of the symmetry of micromagnetic structures and the phenomenological theory of inhomogeneous ME interaction;

*The object of the study* are plates/films of magnetically ordered crystals of cubic, tetragonal, and orthorhombic crystal systems with magnetoelectric properties.

*The subject of the study* is the inhomogeneous magnetoelectric effect in micromagnetic structures.

**Research methods:**

In the course of the study, we performed a group-theoretical analysis, analytical transformations (simplification of phenomenological expressions), and calculations of spatial distributions of electric polarization induced by inhomogeneous ME interactions. To construct a symmetry classification, we enumerated spatial distributions of order parameters of different symmetry and found point groups of magnetic symmetry to which the corresponding distributions are invariant. In order to apply the phenomenological theory of the inhomogeneous ME effect to crystals of a particular symmetry, we found all significantly different linear combinations of invariants of the form $\gamma_{ijkl} P_i M_j \nabla_k M_l$. To demonstrate the application of the phenomenological theory to magnetic vortices, numerical calculations were performed.

**Scientific novelty of the results:**

This thesis is the first to obtain such a symmetry classification of the main micromagnetic structures, including the distribution of structures by their magnetic ("sensitive" to the time reversal operation $1'$) chirality. The organic connection of this classification with the inhomogeneous magnetoelectric effects in them is shown. The inhomogeneous ME effect is qualitatively characterized for each symmetry class. The phenomenological theory of these effects is extended to the case of crystals of cubic, tetragonal, and orthorhombic symmetries. The following new results are obtained:

a) the expressions for the spatial free energy density of inhomogeneous ME interaction for crystals of *each* crystallographic group of the following crystal systems have been found:

  - cubic (*symmetric* groups: $m\bar{3}m1'$, $4321'$, $\bar{4}3m1'$; alternating [39] groups: $m\bar{3}1'$, $231'$)
  - tetragonal (dihedral [39] groups: $4_z/m_z m_x m_{xy} 1'$, $\bar{4}_z 2_x m_{xy} 1'$, $4_z m_x m_y 1'$, $4_z 2_x 2_y 1'$; cyclic [39] groups: $4_z/m_z 1'$, $\bar{4}_z 1'$, $4_z 1'$)
  - orthorhombic (dihedral groups: $mmm1'$, $mm21'$, $2221'$);

b) the groups of magnetic symmetry (including 0° DW) of magnetic DWs (64 classes) and the qualitative characterization of the spatial distribution of magnetization; the inhomogeneous ME effect for each symmetry group is qualitatively described; the magnetic chirality of each DW symmetry group is determined;

c) the groups of magnetic symmetry of BL (48 classes) in the volume of arbitrary magnetic DWs and the qualitative characterization of the structure of the spatial distribution of magnetization; the qualitative characterization of the inhomogeneous ME effect for each symmetry class; the magnetic chirality of each BL symmetry class;

d) classes of magnetic symmetry of BPs (48 classes) in the volume of arbitrary BLs and qualitative characterization of the structure of the spatial distribution of magnetization; qualitative characterization of the inhomogeneous ME effect for each symmetry class; magnetic chirality of each BP symmetry class;
e) the groups of magnetic symmetry of the vortex magnetization distribution in the plate/film were found;

**Practical significance of the results:**

Memory elements and other spintronic devices made on the basis of micromagnetic structures (inventions [22-26], including patents [22,23,26]) can be improved by introducing ME control methods using inhomogeneous ME effects. Improvement is understood as the transition to non-volatile electrostatic control, which qualitatively expands the technological boundaries in the sense of further miniaturization of electronic or spintronic device elements, and increase in operating frequencies. The proposed symmetry classification and phenomenological theory can be used to develop such devices.

The results presented in this thesis can be used to develop methods and interpret the results in experimental studies conducted at the Institute of Magnetism of the National Academy of Sciences of Ukraine, Taras Shevchenko National University of Kyiv, University of Groningen (Netherlands), University of Cologne (Germany), IFW Dresden, and others.

**Personal contribution of the applicant:**

All scientific results were obtained directly by the author or with his decisive participation both in the formulation of tasks and at the stage of theoretical research, interpretation of the results and preparation of works for publication [40-69]. The main part of the publications was prepared with the help of Prof. V. F. Kovalenko and Ph.D. (Physical and Mathematical Sciences) O. V. Tychko. Publications [40-46, 62, 64] were made by the author alone.

**Testing the results:**

Research results have been presented at conferences:

1. VI International Young Scientists' Conference on Applied Physics, Kyiv, Ukraine, 2006, author's reports [60,61].
2. II International Conference "Electronics and Applied Physics", Kyiv, Ukraine, 2006, author's reports [58,59].
3. III International Conference "Electronics and Applied Physics", Kyiv, Ukraine, 2007, author's reports [55,56].
4. International Conference "Functional Materials", Partenit, Crimea, Ukraine, 2007, author's reports [67-69].
5. IV International Conference "Electronics and Applied Physics", Kyiv, Ukraine, 2008, author's report [54].
6. International Scientific Workshop Oxide Materials for Electronic Engineering - fabrication, properties and application (OMEE-2009), Lviv, Ukraine, 2009, author's report [66].
7. International Conference "Functional Materials", Partenit, Crimea, Ukraine, 2009, author's report [65].
8. The 11th Europhysical Conference on Defects in Insulating Materials - EURODIM, Pecs, Hungary, 2010, author's report [64].
9. International Conference "Functional Materials", Partenit, Crimea, Ukraine, 2011, author's reports [62,63].

The main publications of the applicant were cited in 2011-2012 by various researchers in such journals as Journal of Magnetism and Magnetic Materials, Europhysics Letters, Physical Review B - Condensed Matter and Materials Physics, Chemical Science and Solid State Physics.

**Publications:**

On the topic of the thesis, 30 scientific papers were published, including 14 articles in specialized scientific journals [40-53] and 16 conference abstracts [54-69].

**Structure and scope of work:**

The thesis consists of an introduction, four chapters, general conclusions, and a list of references that includes 125 references. The total volume of the thesis is 114 pages, 27 tables and 4 figures.

# SECTION 1

# BIBLIOGRAPHY REVIEW AND TOPICALITY

## 1.1. Multipheroics of type I-II

According to Maxwell's equations, there is a fundamental connection between electric and magnetic interactions. To apply this statement to modern condensed matter physics, it is necessary to consider this statement taking into account relativistic and quantum laws. However, the origin of the connection remains the same.

The assumption of the simultaneous existence of magnetic and electric ordering in a crystal was first made by P. Curie [70] on the basis of a symmetry approach. During experimental studies, for the first time in 1961, a sample of a polycrystal $Pb(Fe_{2/3}W_{1/3})O_3$, which was both a ferroelectric and an antiferromagnetic, was obtained in 1961 [4,71].

According to the terminology that dominates in the current bibliography, multiferroics are materials in which the spatial ordering of magnetization $\boldsymbol{M}$ (and/or antiferromagnetism vector(s) $\boldsymbol{L}$) and electric polarization $\boldsymbol{P}$ coexist in a state of thermodynamic equilibrium in the absence of external fields [72]. In the future, an arbitrary ($\boldsymbol{M}_i$ or $\boldsymbol{L}_k$, where the indices correspond to the presence of sublattices) magnetic order parameter will be denoted as $\boldsymbol{N}$ which must satisfy the condition of time reversal:

$$1' \cdot \boldsymbol{N} = -\boldsymbol{N} \tag{1.1}$$

Multiferroics of the first type are materials that, under the same conditions (fixed external temperature and pressure, absence of external electromagnetic fields), exhibit ferromagnetic and ferroelectric properties simultaneously, which have independent origins and exist independently, i.e., mutual influence $\boldsymbol{P}(r)$ and $N(r)$ are not the cause of ordering. Such coexistence is quite rare [73,74]. In these materials, electrical ordering usually occurs at higher temperatures $T_P$ than the Curie or Néel point ($T_M$) of magnetic ordering. The most common [74] multiferroic of the first type in research and applications is bismuth ferrite $BiFeO_3$whose ME properties are manifested at room temperatures ($T_P=1100\,K$, $T_M=643\,K$). Another example is. $YMnO_3$, whose magnetically ordered phase occurs only at liquid nitrogen temperatures [74] ($T_P=914\,K$, $T_M=76\,K$).

Multiferroics of the second type are materials in which ferroelectric properties arise due to magnetic ordering, i.e., due to ME coupling. Therefore, from the point of view of studying and applying ME effects and interactions, this type of multiferroics deserves the main attention.

Even without analyzing the microscopic mechanism, it turns out to be possible to predict possible ME interactions in arbitrary crystals. The corresponding symmetry classification is developed for arbitrary crystals in the case of interactions between spatially homogeneous orders: $\boldsymbol{P}(r)=\boldsymbol{P}_0$ , and $\boldsymbol{N}(r)=\boldsymbol{N}_0$ [6]. The mechanism of this interaction usually consists in exchange stiction. A classic example of a material in which the linear ME effect is observed is $Cr_2O_3$.

Inhomogeneous ME interaction of spatial distributions $\boldsymbol{P}(r)\neq\boldsymbol{P}_0=const$ and $\boldsymbol{N}(r)\neq\boldsymbol{N}_0=const$ appears among others in spiral multiferroics of the second type. Spiral (cycloidal or helicoidal) magnetic ordering is only a special case of micromagnetic structures. The inhomogeneous ME effect was observed in the following materials: $ZnCr_2Se_4$ [75], $TbMnO_3$ [76], $RMnO_3$ ($R=Dy,Tb$) [77,78], $MnWO_4$ [79], $BiFeO_3$ [80], $Cr_2BeO_4$ [81], $YMn_2O_5$ [82], $HoMnO_3$ [83]. As shown by V.G. Baryakhtar [7] and I.E. Dzyaloshinsky [35], similar inhomogeneous ME interactions are expected not only in spiral multiferroics but also in any magnetically ordered media. This was experimentally shown for the film $(BiLu)_3(FeGa)_5O_{12}$ [31]. Thus, the separation of a separate group of materials with this interaction (inhomogeneous ME interaction) is rather relative and is related to the magnitude of inhomogeneous ME effects, i.e., to the possibility of their observation in modern experiments in any magnetically ordered material.

Composite multiferroics, which are a combination of ferroelectric and magnetically ordered materials, are studied separately [84]. In this case, ME interactions are carried out through the mutual deformation of these materials. Stronger ME coupling, respectively, is achieved in the case of combinations of ferromagnetic and piezoelectric materials.

## 1.2. History of research on inhomogeneous magnetoelectric interactions

### *1.2.1. Development of phenomenological theory*

The form of the free energy density of an inhomogeneous ME interaction can be obtained from general symmetry considerations [85,86]. Such a contribution to the free energy density has the form [9] (hereinafter, Einstein notation is used):

$$F_{ME}(\mathbf{r})=\iint D_{ijn}(\mathbf{r}-\mathbf{r}',\mathbf{r}-\mathbf{r}'')P_i(\mathbf{r})M_j(\mathbf{r}')M_n(\mathbf{r}'')d^3\mathbf{r}'d^3\mathbf{r}'', \qquad (1.2)$$

where integration is performed over the crystal volume. Expression (1.2) describes local (interaction radius $a$ is of the order of the stable lattice) and nonlocal (interaction radius is of the order of the material sample size) interactions. The simplest case of the latter is the ME interaction due to magnetostriction and electrostriction (direct or inverse effects, respectively). An indirect verification of this scheme has been made experimentally [87-89]. The theory and comparison with the experiment in the case of an epitaxial ferroelectric and magnetic nanocomposite film were proposed in [89].

The expansion of (1.2) for a small interaction radius gives the free energy of a local inhomogeneous ME interaction [7]:

$$F_{ME}=\gamma_{ijkl}P_iM_j\nabla_kM_l \qquad (1.3)$$

The structure of the tensor $\hat{\gamma}$ and, accordingly, the entire expression (1.3) is determined by the magnetic point group (crystallographic point group) of the crystal in the simultaneous paramagnetic and paraelectric phase $G_P$ i.e., (1.3) must be an invariant of this group. Local interactions of the type (1.3) will be considered in this paper.

The microscopic origin of the various components in the sum (1.3) was classified by V. G. Baryakhtar [9] as: exchange, exchange-relativistic, and relativistic ME interaction. The most significant contribution to the inhomogeneous ME interaction is made by the exchange component:

$$F_1 = \lambda_{ik} P_i \nabla_k \boldsymbol{M}^2, \tag{1.4}$$

where the tensor $\hat{\lambda}$ (and all subsequent tensors) are obtained from $\hat{\gamma}$. The next most characteristic component is the exchange-relativistic component, which has the form of the Lifshitz invariant:

$$F_2 = d_{ijkl} P_i \left( M_j \nabla_k M_l - M_l \nabla_k M_j \right), \tag{1.5}$$

where $\hat{d}$ is the antisymmetric tensor with respect to the permutation of indices $j$ and $l$. The symmetric relativistic invariant has the smallest characteristic value [9]:

$$F_3 = \beta_{ijkl} P_i \nabla_k \left( M_j M_l \right), \tag{1.6}$$

where $\hat{\beta}$ is the symmetric tensor with respect to the permutation of the indices $j$ and $l$.

### *1.2.2. Overview of microscopic mechanisms*

Experimental observation of the inhomogeneous ME interaction has been made repeatedly. The symmetry analysis allows this interaction in an arbitrary magnetically ordered medium. Nevertheless, the microscopic mechanism of this interaction still requires attention.

Components of the type (1.4) affect the magnetization and/or electric polarization distribution only when the magnetization modulus becomes spatially modulated. A concrete example is the acentric [82] spin density wave, which was shown experimentally [82] for $YMn_2O_5$.

The microscopic mechanism responsible for the contribution to the free energy of the form (1.5) is relatively well studied. The microscopic nature of the exchange-relativistic component of the inhomogeneous ME interaction was proposed on the basis of the spin supercurrent model. Moreover, the expression turns out to be identical (with only different interaction constants $d$) for the superexchange and double exchange interactions [90] (the equilibrium electric polarization of the atomic triad is given instead of the energy expression):

$$\boldsymbol{P}_{ME} = -d\boldsymbol{e}_{12} \times [\boldsymbol{S}_1 \times \boldsymbol{S}_2], \quad (1.7)$$

where $\boldsymbol{S}_1$ and $\boldsymbol{S}_2$ - are the spins of transition metal ions (M) in the triad $M-O-M$ (bonding through the oxygen ion). The unit vector $\boldsymbol{e}_{12}$ is a polar t-even (relative to the time reversal operation $1'$) vector directed from one metal ion to another. It has been shown that it is the Dzyaloshinskii-Moriya interaction that polarizes electron orbitals without analyzing the degrees of freedom of the crystal lattice [90]. At the discovery of this interaction [91-93], ME effects were simply not considered, since the purpose of the relevant studies was the nature of weak ferromagnetism and chiral ordering of magnets. A similar mechanism has been studied in detail for the case of superexchange interaction [90]. A more general (in the sense of including the degrees of freedom of the crystal lattice) theoretical analysis of the microscopic mechanism was presented in [94], where it was shown that the interaction of inhomogeneous magnetization and polarization is implemented through the Dzyaloshinskii-Moriya interaction for certain spin arrangements (a symmetry criterion similar to the microscopic symmetry criteria for weak ferromagnetism in antiferromagnetism). It was also shown there that the effect of "bond bending" of transition metal ions that occurs in this case makes the main contribution to the ferroelectric ordering. A similar theory was proposed in the case of triads $Mn-O-Mn$ in the multiferroics of the perovskite structure $RMnO_3$, $(R=Gd,Tb,Dy)$ [95]. Similarly, the Dzyaloshinskii-Moriya interaction induces the ferroelectric phase through lattice deformations and stabilizes the magnetic structure at low temperatures [95]. The symmetric criteria for the existence of the Dzyaloshinskii-Moriya interaction were formulated in a fundamental study [91-93], which emphasizes the importance of group theory methods in the study of inhomogeneous ME interactions. At the same time, the ME version of the Dzyaloshinskii-Moriya interaction described here requires a special symmetry analysis (the symmetry should additionally allow the existence of a polar t-even vector, i.e., polarization), which was done in [93].

The microscopic spin-dependent polarization can be induced not only through the Dzyaloshinskii-Moriya interaction. A more general form is given by [9,93]:

$$\boldsymbol{P}_{\boldsymbol{ME}}=\lambda\left(\boldsymbol{S}_1\boldsymbol{S}_2\right)+\hat{d}\left[\boldsymbol{S}_1\times\boldsymbol{S}_2\right]+\boldsymbol{S}_1\hat{\Gamma}\,\boldsymbol{S}_2, \tag{1.8}$$

Studies [90,94] describe the second term in (1.8). All components of (1.8) correspond to macroscopic expressions (1.4-1.6) [9]. The microscopic nature of the local inhomogeneous ME interaction can also have a magnetostrictive nature [9]. Also, the Ruderman-Kittel-Kasuya-Yosida interaction can similarly produce electric polarization. As was shown in [96], the anisotropic term of the exchange interaction, which is linear with respect to the spin-orbital interaction, can affect the interaction. Another mechanism is the extended version of the double exchange model including strongly hybridized oxygen's orbitals [97]. A separate microscopic nature of the interaction of electric polarization with spin ordering is found in ferromagnetic Mott insulators, as shown in [98].

In general, the macroscopic symmetry analysis automatically takes into account any microscopic mechanism of inhomogeneous ME interaction – both known and yet to be discovered ones [99].

## 1.3. Development of ideas about magnetic symmetry of micromagnetic structures

Depending on the type of domains separated by the DW, magnetic, ferroelectric, and ferroelectric DWs are distinguished [19].

Establishing the magnetic symmetry of a DW allows us to characterize its structure and possible changes; the corresponding approach was described in the fundamental work [37].

Since DWs can be considered as flat layers, their symmetry is described by 528 magnetic groups of layer symmetry [100,101]. To determine the physical properties at the macroscopic level, it is enough to proceed to the continuum averaging of the microscopic characteristics of the magnetic and electronic structures, which transforms these groups into magnetic point groups [102] and makes these structures invariant with respect to translations in the crystal lattice. At the same time, the continuous translation operation is traditionally considered as the identity operation, since it is present in all groups and does not affect the systematization of the structure. It has been shown [103] that, in the most general formulation of the problem of the structure of a plane DW, there are 125 magnetic point groups. At the same time, as a simple criterion, the requirement of pyroelectric and/or gyromagnetic character of these groups was considered, which allows spontaneous $\boldsymbol{P}$ and/or $\boldsymbol{M}$ in the volume of DW, respectively [104-106]. These criteria, respectively, only allow us to predict homogeneous distributions or even components in the spatial distributions $\boldsymbol{P}(\boldsymbol{r})$ and/or $\boldsymbol{M}(\boldsymbol{r})$. To fully establish the nature of the DW structure, this approach needs to be refined.

For magnetic DWs, there are 42 groups in the case of 180° DWs [37] and 10 groups in the case of DWs with non-collinear domain magnetizations [107]. For magnetic 0° DWs, the symmetry classification has not yet been constructed.

Comparison of the symmetry of the medium and the symmetry of the DW allows us to obtain DW states with the same energy [38]. The corresponding DW regions are the boundary conditions for BLs. Symmetry classification of BLs inside magnetic 180° DWs, which includes 32 magnetic point symmetry groups, has been constructed [38]. In the above studies of the symmetry of micromagnetic structures, their chirality was not considered.

Symmetry criteria for the occurrence of localized magnetic solitons were formulated in [108]. It was found that magnetically ordered crystals should belong to the crystallographic groups $C_n$, $C_{nv}$, $D_n$, $D_{2d}$ or $S_4$, where $n=3,4,6$. At the same time, at certain values of the external field, not vortex structures but a set of non-interacting 0° DWs can emerge, which emphasizes the common nature of different solutions of the soliton type.

### 1.4. Concept of chirality of micromagnetic structures

Due to the insufficient theoretical basis of group-theoretical methods as applied to micromagnetic structures, there are problems (lack of convention) with the use of the concept of chirality (enantiomorphism) in the study of micromagnetic structures. At scientific conferences, there is a lack of uniform terminology in magnetism regarding chirality. At the same time, this problem is quite fundamental in the sense of interaction between the physics of magnetism and other areas of research since chirality is an interdisciplinary concept. Examples include stereochemistry, magneto-chiral dichroism studies, homochirality of living systems, etc.

The term "chirality" was first introduced by Kelvin [109]. This property is widely used in all fields of applied and theoretical physics. In the case of magnetic symmetry, the problem of chirality definition was formulated in the concept of "complete symmetry" by Zheludev [110] and the studies of Barron [111]. The latter assumes that in the case of systems with motion (including magnetically ordered media), different types of chirality exist with respect to the time direction parity. Thus, *t-even* and *t-odd* chirality were separated [112]. The difference between Barron and Kelvin's definitions of chirality is in the emergence of these two types of chiral objects. The t-odd chirality was called false chirality in contrast to t-even (true) chirality [113]. Such "truth" and "falsity" is a purely subjective choice of terminology.

In the study of magnetic vortices, the definition of chirality is used as a continuous parameter [114]. In the case of frustrated magnetics, additional definitions of chirality arise: vector [115] and scalar [116] chirality. These definitions do not coincide with the Kelvin definition. The vector chirality is used to study non-collinear magnetic ordering (including cycloidal). In this case, the rotation of the cycloid around the second-order symmetry axis turns the "right" (according to the vector definition) cycloid into the "left" one.

The definition of chirality in a magnetically ordered medium can be aligned with the geometric Kelvin definition to use the same formal language in different disciplines. For example, stereochemistry and condensed matter physics are related through the synthesis of materials.

This correspondence is achieved by using the Barron chirality [112,113], which can be defined according to the following formal scheme. A chiral point group of magnetic symmetry should not contain symmetry operations of the type $\bar{n}$ where $n$ - is an integer, i.e., the symmetry group should not contain any inverse spatial rotations (without time reversal) including mirror reflections $\bar{2} \equiv m$. In this case, if the group contains operations of type $\bar{n}'$ then the chirality is t-odd. Otherwise, the chirality is t-even.

# SECTION 2

# SYMMETRY OF MICROMAGNETIC STRUCTURES

## 2.1. Symmetry classification of magnetic domain walls

Let us select the coordinate system $XYZ$ associated with the DW plane (the Z-axis is directed along the normal of this plane). The spatial distribution of electric polarization is as follows:

$$\boldsymbol{P}(z)=\boldsymbol{e}_x P_x(z)+\boldsymbol{e}_y P_y(z)+\boldsymbol{e}_z P_z(z) \tag{2.1}$$

The magnetic point symmetry group consists of two types of operations. Operations of the first type $g^{(1)}$ do not change the spatial coordinate $z$. If the operation $g^{(1)}$ is a rotation around the symmetry axis of $n$-th ($n>1$) order $n_z$, then the symmetry imposes a restriction:

$$P_x(z)=P_y(z)=0 \tag{2.2}$$

Operations of the second type $g^{(2)}$ have a different property:

$$g^{(2)} z=-z \tag{2.3}$$

Operations $g^{(1)}$ and $g^{(2)}$ can leave only even (S) and odd (A) component dependencies $P_i(z)$ as invariants respectively, where $i=x,y,z$. If the magnetic point symmetry group of the DW is not polar, then it holds:

$$\boldsymbol{P}(z)=-\boldsymbol{P}(-z) \tag{2.4}$$

In this case, the component $P_z(z)$ is necessarily nonzero. It follows from this criterion that 125 magnetic point groups allow for the existence of spontaneous $P$ in the DW volume.

There is no magnetization in the DW volume ($\boldsymbol{M}(z)=0$) if the operations $g^{(1)}$ are: time reversal operation $1'$ and/or rotation around the symmetry axis $l_z'$ ($l>2$) and/or rotation about the symmetry axis of the $n$-th order $n_z$ ($n>1$), which exist simultaneously with the mirroring operation $m_\perp$ in the perpendicular orientation relative to the DW plane. In the volume of a DW with other point symmetry groups that are not pyromagnetic, there is a distribution of magnetization of the form:

$$\boldsymbol{M}(z) = -\boldsymbol{M}(-z) \tag{2.5}$$

Only 64 groups out of 125 point groups allow the existence of a nonzero function $\boldsymbol{M}(z)$. These groups correspond to magnetic DWs.

There are 64 point symmetry groups (we number them by indices $1 \le k \le 64$) of magnetic DWs, including 42 groups of 180° DWs, 42 groups of 0° DWs, and 10 groups of DWs with non-collinear magnetizations of neighboring domains.

Since $m\bar{3}m$ crystal does not contain symmetric axes of rotation of the 6th order (including inversion axes), then, limiting ourselves to a cubic crystal, there are 52 magnetic point groups. These groups are listed in Tables 2.1-2.4. The notation (A,S) corresponds to the sum of an even and an odd function of the spatial coordinate.

Chiral point groups of magnetic symmetry of magnetic DWs do not contain symmetry operations of the type $\bar{n}$. For the case of t-even chirality, an additional requirement is the absence of operations $\bar{n}'$ (Table 2.1). If the chiral group contains such operations, then the DW is characterized by t-odd chirality (Table 2.2).

**Table 2.1**. Point groups of magnetic symmetry of magnetic DWs with t-even chirality

| $k$ | **Group** | $M_x(z)$ | $M_y(z)$ | $M_z(z)$ | $P_x(z)$ | $P_y(z)$ | $P_z(z)$ |
|---|---|---|---|---|---|---|---|
| 7 | $2_x^{'}2_y2_z^{'}$ | A | S | 0 | 0 | 0 | A |
| 8 | $2_z^{'}$ | A,S | A,S | 0 | 0 | 0 | A,S |
| 10 | $2_x^{'}$ | A | S | S | S | A | A |
| 13 | $2_y$ | A | S | A | A | S | A |
| 16 | 1 | A,S | A,S | A,S | A,S | A,S | A,S |
| 19 | $2_z$ | 0 | 0 | A,S | 0 | 0 | A,S |
| 21 | $2_x2_y2_z$ | 0 | 0 | A | 0 | 0 | A |
| 24 | $3_z$ | 0 | 0 | A,S | 0 | 0 | A,S |
| 27 | $3_z2_x$ | 0 | 0 | A | 0 | 0 | A |
| 30 | $4_z$ | 0 | 0 | A,S | 0 | 0 | A,S |
| 33 | $4_z2_x2_{xy}$ | 0 | 0 | A | 0 | 0 | A |
| 50 | $2_z2'_x2'_y$ | 0 | 0 | S | 0 | 0 | A |
| 53 | $3_z2_x^{'}$ | 0 | 0 | S | 0 | 0 | A |
| 57 | $4_z2_x^{'}2_y^{'}$ | 0 | 0 | S | 0 | 0 | A |

The presence of at least one symmetry operation of type $\bar{n}$ is a criterion for the absence of enantiomorphism. Achiral magnetic DWs are described by the largest set of point groups of magnetic symmetry (Tables 2.3-2.4).

**Table 2.2**. Point groups of magnetic symmetry of magnetic DWs with t-odd chirality

| $k$ | **Group** | $M_x(z)$ | $M_y(z)$ | $M_z(z)$ | $P_x(z)$ | $P_y(z)$ | $P_z(z)$ |
|---|---|---|---|---|---|---|---|
| 12 | $m_x'$ | 0 | A,S | A,S | 0 | A,S | A,S |
| 14 | $2_y/m_y'$ | A | 0 | A | A | 0 | A |
| 15 | $\bar{1}'$ | A | A | A | A | A | A |
| 17 | $m_x'm_z'2_y$ | 0 | S | A | 0 | S | A |
| 18 | $m_z'$ | S | S | A | S | S | A |
| 20 | $2_z/m_z'$ | 0 | 0 | A | 0 | 0 | A |
| 22 | $m_x'm_y'2_z$ | 0 | 0 | A,S | 0 | 0 | A,S |
| 23 | $m_x'm_y'm_z'$ | 0 | 0 | A | 0 | 0 | A |
| 26 | $3_zm_x'$ | 0 | 0 | A,S | 0 | 0 | A,S |
| 29 | $\bar{3}'_zm_x'$ | 0 | 0 | A | 0 | 0 | A |
| 31 | $4_z/m_z'$ | 0 | 0 | A | 0 | 0 | A |
| 32 | $4_zm_x'm_{xy}'$ | 0 | 0 | A,S | 0 | 0 | A,S |
| 34 | $4_z/m_z'm_x'm_{xy}'$ | 0 | 0 | A | 0 | 0 | A |
| 35 | $\bar{4}_z'$ | 0 | 0 | A | 0 | 0 | A |
| 36 | $\bar{4}_z'2_xm_{xy}'$ | 0 | 0 | A | 0 | 0 | A |
| 42 | $\bar{3}_z'$ | 0 | 0 | A | 0 | 0 | A |

**Table 2.3**. Point groups of magnetic symmetry of magnetic achiral DWs of rotary type

| $k$ | **Group** | $M_x(z)$ | $M_y(z)$ | $M_z(z)$ | $P_x(z)$ | $P_y(z)$ | $P_z(z)$ |
|---|---|---|---|---|---|---|---|
| 9 | $m_z m'_y 2'_x$ | A | 0 | S | S | 0 | A |
| 11 | $m_z$ | A | A | S | S | S | A |
| 46 | $2'_z/m'_z$ | S | S | 0 | 0 | 0 | A |
| 47 | $2'_x/m'_x$ | 0 | S | S | 0 | A | A |
| 48 | $\bar{1}$ | S | S | S | A | A | A |

A point group of magnetic symmetry requiring two zero components of $\boldsymbol{M}$ or all odd components corresponds to the cases $|\boldsymbol{M}| \neq const$. Such DWs exist in the vicinity of phase transition points [37]. Accordingly, we distinguish between "pulsating" and "rotating" DWs (as suggested in [37]). The corresponding systematization is presented only for achiral structures (Tables 2.3-2.4).

The point group of magnetic symmetry of DW is a subgroup of the group $G_P$ of the magnetic symmetry group of the magnetic crystal sample in the paramagnetic phase [37]. Taking into account the influence of free surfaces of the sample on the micromagnetic structure, the group $G_P$ should be considered as the intersection of the point group of magnetic symmetry $G_S$ of the shape of the sample's surfaces with the point group of magnetic symmetry of an unbounded sample:

$$G_P^{\infty} = G \otimes (1, 1'), \tag{2.6}$$

where, $G$ - crystallographic class. For the cases of a plate or film, $G_S$ is given by:

$$G_S = \infty / mmm1' \tag{2.7}$$

Thus, according to the Curie principle, the following equality holds:

$$G_P = \infty_{[hkl]} / mmm1' \cap m\bar{3}m1', \tag{2.8}$$

where $(hkl)$ - is the crystallographic orientation of the plates/films.

Group-theoretical methods allow one to determine the degeneracy of DWs [53], i.e., to find DWs with the identical energy but different structures. The degree of degeneracy $q_W$ is defined as the ratio of group orders:

$$q_W = |G_P| / |G_k|, \qquad (2.9)$$

where $G_k$ is a point symmetry group of the DW. If one chooses a certain pair of neighboring domains, one can distinguish between the following types of degeneracy: degeneracy of a DW under fixed boundary conditions:

$$q'_W = |G_B| / |G_k| \qquad (2.10)$$

and degeneracy of the boundary conditions themselves:

$$q_B = |G_P| / |G_B| \qquad (2.11)$$

Here, the group $G_B$ - is a point group of magnetic symmetry of boundary conditions (i.e., a combination of spatially separated magnetic domains) [37]. A group $G_B$ is a subgroup of $G_P$ that is determined by the orientation of the DWs, the type of DWs, and the directions of the domains' magnetizations relative to the sample surface.

**Table 2.4**. Point groups of magnetic symmetry of magnetic achiral DWs of pulsating type

| $k$ | **Group** | $M_x(z)$ | $M_y(z)$ | $M_z(z)$ | $P_x(z)$ | $P_y(z)$ | $P_z(z)$ |
|---|---|---|---|---|---|---|---|
| 1 | $m_x m_z m_y'$ | A | 0 | 0 | 0 | 0 | A |
| 2 | $m'_y m_x 2'_z$ | A,S | 0 | 0 | 0 | 0 | A,S |
| 3 | $m_x m_z 2_y$ | A | 0 | 0 | 0 | S | A |
| 4 | $2_x'/m_x$ | A | 0 | 0 | 0 | A | A |
| 5 | $2_z'/m_z$ | A | A | 0 | 0 | 0 | A |
| 6 | $m_y$ | A,S | 0 | 0 | 0 | A,S | A,S |
| 43 | $m_y m_x' m_z'$ | 0 | S | 0 | 0 | 0 | A |
| 44 | $m_y m_z' 2_x'$ | 0 | S | 0 | S | 0 | A |
| 45 | $2_y/m_y$ | 0 | S | 0 | A | 0 | A |
| 49 | $2_z/m_z$ | 0 | 0 | S | 0 | 0 | A |
| 51 | $m_z m_x' m_y'$ | 0 | 0 | S | 0 | 0 | A |
| 55 | $\bar{3}_z m_x'$ | 0 | 0 | S | 0 | 0 | A |
| 56 | $4_z/m_z$ | 0 | 0 | S | 0 | 0 | A |
| 58 | $4_z/m_z m_x' m_{xy}'$ | 0 | 0 | S | 0 | 0 | A |
| 59 | $\bar{4}_z$ | 0 | 0 | S | 0 | 0 | A |
| 60 | $\bar{4}_z 2_x' m_{xy}'$ | 0 | 0 | S | 0 | 0 | A |
| 64 | $\bar{3}_z$ | 0 | 0 | S | 0 | 0 | A |

It is important to note that each of the groups $G_k$ is one of the groups $G_B$ and vice versa, their general lists are identical (Tables 2.1-2.4). There are $q_W$ groups and corresponding $q_W$ symmetry operations $g_i^{(l)}$ (members of adjacent groups [37]) in the decomposition:

$$G_P = g_0^{(l)} G_k + g_1^{(l)} G_k + g_2^{(l)} G_k + \ldots + g_{q_W}^{(l)} G_k \,, \qquad (2.12)$$

where $g_0^{(l)} \equiv 1$. These, "lost" operations $g_i^{(l)}$ transform DWs with the same energy into each other.

## 2.2. Symmetry classification of the Bloch lines

One of the key applications of the symmetric DW classification is an ability to predict the degeneracy of DW under any boundary conditions in a crystal with any symmetry.

If one considers the spontaneous mechanism of [38] BL occurrence, the DW degeneration scheme makes it possible to predict the structure of arbitrary [38] BLs. Each combination of two equal-energy DW sections corresponds to a particular BL. The corresponding sections are directly obtained from the expansions of the form (2.12). At the same time, a pair of DW sections determines the boundary conditions for the magnetization distribution in the volume of the BL. The structure of the BL under the same boundary conditions can be different. The symmetric approach seems to be the most rational for systematization of possible BLs.

According to the theory of phase transitions, the ordering of a solid by some physical (non-scalar) quantity is accompanied by a decrease in symmetry to one of the subgroups of the original symmetry group, which in our case means

$$G_k \subseteq G_B \subset G_P, \tag{2.13}$$

Each DW site can be combined with $q_W^{'}-1$ possible sections. Accordingly, the total number of boundary conditions (we denote the symmetry group of boundary conditions as $U_B \subseteq G_B$) of a BL in a given DW is:

$$Q_B = \left(q_W^{'}-1\right) q_W^{'} \tag{2.14}$$

Not all subgroups of $U_B \subseteq G_B$ are allowed, since the symmetry of the BL depends on the symmetry of the DW, i.e., on the group $G_k$ and lost operations [37], which establish the correspondence between the DW sections.

For a given pair of DW sites (given boundary conditions for BLs), there is possible this number of BLs:

$$Q_l^{'} = |U_B| / |U_l|, \tag{2.15}$$

where $U_l$ - is $l$-th point group of magnetic symmetry of the BL. These BLs have identical symmetries of their structures and the same free energies, but the structures themselves are different. The correspondence between the implementation of degenerate BLs is established by a new type of lost symmetry operations $u_i^{(l)}$ in the decomposition:

$$U_B = u_0^{(l)} U_l + u_1^{(l)} U_l + \ldots + u_{Q_l}^{(l)} U_l \, , \qquad (2.16)$$

where $u_0^{(l)} \equiv 1$. Finally, the total number of all possible BLs with the same energy is given by the expression:

$$N_{BL} = Q_B Q_l^{'} \qquad (2.17)$$

In the spontaneous mechanism [38] of BL generation in the domain structure, the symmetry of the DW boundary conditions is reduced or preserved:

$$U_l \subseteq U_B \subseteq G_B \subset G_P, \qquad (2.18)$$

The relation between the point group of magnetic symmetry and the allowed functions $\boldsymbol{M}(\boldsymbol{r})$ and $\boldsymbol{P}(\boldsymbol{r})$ was formulated in [38]. Let us choose a coordinate system associated with the DW plane: the $Z$ axis is normal to the DW plane, the $Y$ axis is directed along the BL. The magnetization distribution is given by:

$$\boldsymbol{M}(z,x) = M_x(z,x)\boldsymbol{e}_x + M_y(z,x)\boldsymbol{e}_y + M_z(z,x)\boldsymbol{e}_z \qquad (2.19)$$

The distribution of electric polarization is set similarly. The algorithm for constructing the symmetry classification (Tables 2.5-2.8) is similar to that used in the DW. Different types of chirality are separately indicated in the presented classification.

A record of the type (A,S)/A under the component $M_i(z,x)$ means that $i$-th component of the vector changes sign (is asymmetric) when $z \to -z$; one does not change sign (is symmetric) when $x \to -x$; and, finally, one changes sign when both coordinates $z$ and $x$ changing sign. Lowercase letters indicate components that are equal to zero in the volume of domains.

**Table 2.5**. Point groups of magnetic symmetry of BL with t-even chirality.

| **Group** | $M_x(z,x)$ | $M_y(z,x)$ | $M_z(z,x)$ | $P_x(z,x)$ | $P_y(z,x)$ | $P_z(z,x)$ |
|---|---|---|---|---|---|---|
| $2'_x 2_y 2'_z$ | (A,S)/A | (S,S)/S | (s,a)/a | (s,a)/a | (a,a)/s | (a,s)/a |
| $2'_y 2_x 2'_z$ | (S,S)/S | (A,S)/A | (a,a)/s | (s,a)/a | (a,a)/s | (a,s)/a |
| $2'_z$ | (F,S)/F | (F,S)/F | (f,a)/f | (f,a)/f | (f,a)/f | (f,s)/f |
| $2'_x$ | (A,F)/F | (S,F)/F | (S,F)/F | (s,f)/f | (a,f)/f | (a,f)/f |
| $2'_y$ | (F,F)/S | (F,F)/A | (F,F)/S | (f,f)/a | (f,f)/s | (f,f)/a |
| $2_y$ | (F,F)/A | (F,F)/S | (F,F)/A | (f,f)/a | (f,f)/s | (f,f)/a |
| $2_x$ | (S,F)/F | (A,F)/F | (A,F)/F | (s,f)/f | (a,f)/f | (a,f)/f |
| 1 | (F,F)/F | (F,F)/F | (F,F)/F | (f,f)/f | (f,f)/f | (f,f)/f |
| $2_z$ | (f,a)/f | (f,a)/f | (F,S)/F | (f,a)/f | (f,a)/f | (f,s)/f |
| $2_x 2_y 2_z$ | (s,a)/a | (a,a)/s | (A,S)/A | (s,a)/a | (a,a)/s | (a,s)/a |
| $2_z 2'_x 2'_y$ | (a,a)/s | (s,a)/a | (S,S)/S | (s,a)/a | (a,a)/s | (a,s)/a |

**Table 2.6.** Point groups of magnetic symmetry of BL with t-odd chirality

| Group | $M_x(z,x)$ | $M_y(z,x)$ | $M_z(z,x)$ | $P_x(z,x)$ | $P_y(z,x)$ | $P_z(z,x)$ |
|---|---|---|---|---|---|---|
| $m_x^{'}$ | (f,a)/f | (F,S)/F | (F,S)/F | (f,a)/f | (f,s)/f | (f,s)/f |
| $m_y^{'}$ | (F,F)/F | (0) | (F,F)/F | (f,f)/f | (0) | (f,f)/f |
| $2_y/m_y{}'$ | (F,F)/A | (0) | (F,F)/A | (f,f)/a | (0) | (f,f)/a |
| $2_x/m_x{}'$ | (s,a)/a | (A,S)/A | (A,S)/A | (s,a)/a | (a,s)/a | (a,s)/a |
| $\bar{1}'$ | (F,F)/A | (F,F)/A | (F,F)/A | (f,f)/a | (f,f)/a | (f,f)/a |
| $m_x^{'}m_z^{'}2_y$ | (s,a)/a | (S,S)/S | (A,S)/A | (s,a)/a | (s,s)/s | (a,s)/a |
| $m_y^{'}m_z^{'}2_x$ | (S,F)/F | (0) | (A,F)/F | (s,f)/f | (0) | (a,f)/f |
| $m_z^{'}$ | (S,F)/F | (S,F)/F | (A,F)/F | (s,f)/f | (s,f)/f | (a,f)/f |
| $2_z/m_z^{'}$ | (s,a)/a | (s,a)/a | (A,S)/A | (s,a)/a | (s,a)/a | (a,s)/a |
| $m_x^{'}m_y^{'}2_z$ | (f,a)/f | (0) | (F,S)/F | (f,a)/f | (0) | (f,s)/f |
| $m_x^{'}m_y^{'}m_z^{'}$ | (s,a)/a | (0) | (A,S)/A | (s,a)/a | (0) | (a,s)/a |

In general, there are 48 point groups of magnetic symmetry of BLs including 22 (11 t-even and 11 t-odd) chiral and 26 achiral groups. This is a symmetry classification of BLs (ferromagnetic, ferroelectric, ferroelectric and mixed types) in planar magnetic DWs in an arbitrary magnetically ordered medium. A subset of this classification was obtained for ferromagnetic DWs in 180° DWs before [38]. Our results are in agreement with [38].

The definition of rotary DWs [37] is also applicable to BLs, where we limit ourselves to achiral BLs. The criterion for rotary achiral BLs is the presence of at least two non-zero magnetization components.

BLs with a non-constant spatial distribution of the magnetization modulus should be considered separately. They correspond to symmetry groups that require two zero magnetization components (Table 2.8).

**Table 2.7.** Point groups of magnetic symmetry of achiral BLs of rotary type.

| **Group** | $M_x(z,x)$ | $M_y(z,x)$ | $M_z(z,x)$ | $P_x(z,x)$ | $P_y(z,x)$ | $P_z(z,x)$ |
|---|---|---|---|---|---|---|
| $m_x m_z m'_y$ | (A,S)/A | (0) | (s,a)/a | (s,a)/a | (0) | (a,s)/a |
| $m'_y m_x 2'_z$ | (F,S)/F | (0) | (f,a)/f | (f,a)/f | (0) | (f,s)/f |
| $m_x m_z 2_y$ | (A,S)/A | (a,a)/s | (s,a)/a | (s,a)/a | (s,s)/s | (a,s)/a |
| $2'_x/m_x$ | (A,S)/A | (s,a)/a | (s,a)/a | (s,a)/a | (a,s)/a | (a,s)/a |
| $2'_z/m_z$ | (A,S)/A | (A,S)/A | (s,a)/a | (s,a)/a | (s,a)/a | (a,s)/a |
| $m_x$ | (F,S)/F | (f,a)/f | (f,a)/f | (f,a)/f | (f,s)/f | (f,s)/f |
| $m_z m'_y 2'_x$ | (A,F)/F | (0) | (S,F)/F | (s,f)/f | (0) | (a,f)/f |
| $m_z m'_x 2'_y$ | (a,a)/s | (A,S)/A | (S,S)/S | (s,a)/a | (s,s)/s | (a,s)/a |
| $m_z$ | (A,F)/F | (A,F)/F | (S,F)/F | (s,f)/f | (s,f)/f | (a,f)/f |
| $m_x m'_y m'_z$ | (S,S)/S | (0) | (a,a)/s | (s,a)/a | (0) | (a,s)/a |
| $m_x m'_z 2'_y$ | (S,S)/S | (s,a)/a | (a,a)/s | (s,a)/a | (s,s)/s | (a,s)/a |
| $2_x/m_x$ | (S,S)/S | (a,a)/s | (a,a)/s | (s,a)/a | (a,s)/a | (a,s)/a |
| $2'_z/m'_z$ | (S,S)/S | (S,S)/S | (a,a)/s | (s,a)/a | (s,a)/a | (a,s)/a |
| $2'_x/m'_x$ | (a,a)/s | (S,S)/S | (S,S)/S | (s,a)/a | (a,s)/a | (a,s)/a |
| $2'_y/m'_y$ | (F,F)/S | (0) | (F,F)/S | (f,f)/a | (0) | (f,f)/a |
| $\bar{1}$ | (F,F)/S | (F,F)/S | (F,F)/S | (f,f)/a | (f,f)/a | (f,f)/a |
| $2_z/m_z$ | (a,a)/s | (a,a)/s | (S,S)/S | (s,a)/a | (s,a)/a | (a,s)/a |
| $m_z m'_x m'_y$ | (a,a)/s | (0) | (S,S)/S | (s,a)/a | (0) | (a,s)/a |

First of all, the presented classification describes directly the symmetry of BLs (group $U_l$). Additionally, tables 2.5-2.8 give a list of all possible groups $U_B$ (as in the case of DWs, these lists coincide). In the latter case, the functional dependencies describe the relationship between the boundary conditions, i.e., between the vectors $M(x \to \pm\infty)$ and $P(x \to \pm\infty)$which are separated spatially. This allows us to determine the type of degeneration of the BL, similar to the way it was done for the degeneration of the DW. To do this, it is necessary to compare the groups $U_l$ and $U_B$. If the magnetization components are different for the groups $U_l$ ($U_l \subseteq U_B$) and $U_B$ then the degeneration is ferromagnetic. If at least one electric polarization component is of different types in the case of these groups, then the degeneration is ferroelectric. These two types can be combined. The criteria can be formulated more formally.

**Table 2.8.** Point groups of magnetic symmetry of achiral BLs with one magnetization component.

| Group | $M_x(z,x)$ | $M_y(z,x)$ | $M_z(z,x)$ | $P_x(z,x)$ | $P_y(z,x)$ | $P_z(z,x)$ |
|---|---|---|---|---|---|---|
| $m_y m_z m_x'$ | (0) | (A,S)/A | (0) | (s,a)/a | (0) | (a,s)/a |
| $m'_x m_y 2'_z$ | (0) | (F,S)/F | (0) | (f,a)/f | (0) | (f,s)/f |
| $m_y m_z 2_x$ | (0) | (A,F)/F | (0) | (s,f)/f | (0) | (a,f)/f |
| $2_y'/m_y$ | (0) | (F,F)/A | (0) | (f,f)/a | (0) | (f,f)/a |
| $m_y$ | (0) | (F,F)/F | (0) | (f,f)/f | (0) | (f,f)/f |
| $m_y m_x' m_z'$ | (0) | (S,S)/S | (0) | (s,a)/a | (0) | (a,s)/a |
| $m_y m_z' 2_x'$ | (0) | (S,F)/F | (0) | (s,f)/f | (0) | (a,f)/f |
| $2_y/m_y$ | (0) | (F,F)/S | (0) | (f,f)/a | (0) | (f,f)/a |

If the functions $\boldsymbol{M}(\boldsymbol{r})$ and/or $\boldsymbol{P}(\boldsymbol{r})$ are not invariants of at least one lost symmetry operation $u_i^{(l)}$ of the expansion (2.16), then the degeneration is ferromagnetic and/or ferroelectric, respectively. The ferroelectric degeneration can be defined similarly, but is not studied here.

The t-even chiral BLs can have the same zero magnetization and polarization components $M_y(z,x)$ and polarization $P_y(z,x)$ (Table 2.6-2.7). Achiral BLs can also have zero magnetization components:

$$M_x(z,x)=M_z(z,x)=0 \tag{2.20}$$

and polarization $P_y(z,x)=0$ simultaneously (Table 2.8). BLs with t-odd chirality have qualitatively identical spatial distributions $\boldsymbol{M}(z,x)$ and $\boldsymbol{P}(z,x)$ (Table 2.6). This property remains true for all micromagnetic structures.

## 2.3. Symmetry classification of Bloch points

For the case of a spontaneous mechanism of BP emergence in the volume of BLs, the structure of BPs can be described (magnetic point symmetry group $W_p$), based on the symmetry classification of BLs and their degeneration. The degenerate state of a BL corresponds to the possibility of establishing $Q_l^{'}$ equal-energy BL sections as boundary conditions (magnetic point symmetry group $W_B$) for a BP whose structure stabilizes to a state with identical or lower symmetry:

$$W_p \subseteq W_B \subseteq U_B \subseteq G_B \subset G_P \quad (2.21)$$

Due to the geometry of the problem of BP and BL on magnetic planar DWs, all symmetry groups of both BL and BP are subgroups of the group $mmm1'$ symmetry group of the corresponding geometric model (consisting of a plane, a line, a point and the normal to the plane):

$$W_B \subseteq U_B \subset mmm1' \quad (2.22)$$

Each BL site can be combined with $Q_l^{'} - 1$ possible BL sections with identical structure and free energy. Accordingly, the total number of boundary conditions of the BP that arose on the basis of a given BL is given by:

$$\widetilde{Q}_B = \left(Q_l^{'} - 1\right) Q_l^{'} \quad (2.23)$$

Not all subgroups of this relation:

$$W_B \subseteq U_B \subseteq G_B \quad (2.24)$$

are allowed, since the symmetry of BP depends on the symmetry of BL, i.e., on the group $U_l$ and the corresponding lost operations. Accordingly, the true number of BP boundary conditions is given by expression (2.23):

$$\widetilde{Q}_B \leq |U_B| / |W_B| \quad (2.25)$$

Under fixed boundary conditions, an additional degeneration of the equilibrium state of the BP micromagnetic structure (spatial distribution of magnetization and electric polarization) is possible, which corresponds to the last symmetry series in expression (2.21):

$$W_B = w_0^{(l)} W_p + w_1^{(l)} W_p + \ldots + w_{\tilde{Q}_p'}^{(l)} W_p \ , \qquad (2.26)$$

where $w_0^{(l)} \equiv 1$ and the symbol "$l$" denotes the lost symmetry operations. Finally, the total number of all possible BPs with the same energy is given by the expression:

$$N_{BP} = \tilde{Q}_B \tilde{Q}_p' \qquad (2.27)$$

Let us choose a coordinate system associated with the DW plane: the $Z$ axis is normal to the DW plane, the $Y$ axis is directed along the BL. The magnetization distribution is given as:

$$\boldsymbol{M}(z,x,y) = M_x(z,x,y)\boldsymbol{e}_x + M_y(z,x,y)\boldsymbol{e}_y + M_z(z,x,y)\boldsymbol{e}_z \qquad (2.28)$$

The distribution of electric polarization is defined in a similar way. Different types of magnetic chirality are separately indicated in the presented classification.

Tables 2.9-2.16 use the notation of the types of spatial distributions of magnetization and polarization in the form $(T_z, T_x, T_y)/T_{zx}/T_{xy}/T_{zy}/./T_{zxy}$ for the corresponding types of components of the corresponding order parameter. The ending of this notation is provided only when it is not trivially derived from its beginning. The symbol "$T$" denotes the type of spatial distribution of the component: even ("S", symmetric) and odd ("A", antisymmetric) functions, or the sum of even and odd functions (F) for the corresponding spatial parities: $z \to -z$ (type $T_z$), $x \to -x$ (type $T_x$), $y \to -y$ (type $T_y$) and simultaneously transferring $z \to -z$, $x \to -x$ (type $T_{zx}$), $x \to -x$, $y \to -y$ (type $T_{xy}$), $z \to -z$, $y \to -y$ (type $T_{zy}$), full spatial inversion of coordinates $T_{zxy}$.

**Table 2.9.** Point groups of BP magnetic symmetry with t-even chirality. Groups of the first and second orders.

| **Group** | $M_x$ | $M_y$ | $M_z$ | $P_x$ | $P_y$ | $P_z$ |
|---|---|---|---|---|---|---|
| $2_z'$ | F/S/F | F/S/F | F/A/F | F/A/F | F/A/F | F/S/F |
| $2_x'$ | F/F/A | F/F/S | F/F/S | F/F/S | F/F/A | F/F/A |
| $2_y'$ | S/F/F | A/F/F | S/F/F | A/F/F | S/F/F | A/F/F |
| $2_y$ | A/F/F | S/F/F | A/F/F | A/F/F | S/F/F | A/F/F |
| $2_x$ | F/F/S | F/F/A | F/F/A | F/F/S | F/F/A | F/F/A |
| 1 | F/F/F | F/F/F | F/F/F | F/F/F | F/F/F | F/F/F |
| $2_z$ | F/A/F | F/A/F | F/S/F | F/A/F | F/A/F | F/S/F |

Different symmetry groups of BP are systematized by the order of the symmetry group. A lower order of the group (number of symmetry transformations in its representation) corresponds to a larger number of arbitrary spatial distributions (type *F*), i.e., the requirements imposed by the symmetry are weaker. Thus, in the case of a first-order group (identification operation 1), there are no such requirements (Table 2.9). For the point groups of magnetic symmetry of the second order, only one functional dependence is defined (Tables 2.9, 2.11).

**Table 2.10.** Point groups of BP magnetic symmetry with t-even chirality. Groups of the fourth order.

| **Group** | $M_x$ | $M_y$ | $M_z$ | $P_x$ | $P_y$ | $P_z$ |
|---|---|---|---|---|---|---|
| $2_x^{'}2_y2_z^{'}$ | A/S/A | S/S/S | A/A/S | A/A/S | S/A/A | A/S/A |
| $2_y^{'}2_x2_z^{'}$ | S/S/S | A/S/A | S/A/A | A/A/S | S/A/A | A/S/A |
| $2_x2_y2_z$ | A/A/S | S/A/A | A/S/A | A/A/S | S/A/A | A/S/A |
| $2_z2'_x2'_y$ | S/A/A | A/A/S | S/S/S | A/A/S | S/A/A | A/S/A |

In the case of groups of the fourth order, three functional dependencies are defined (Table 2.12). In the case of groups of the eighth order, all 7 functional dependencies are determined (Table 2.12).

**Table 2.11.** Point groups of BP magnetic symmetry with t-odd chirality. Groups of the second order.

| **Group** | $M_x$ | $M_y$ | $M_z$ | $P_x$ | $P_y$ | $P_z$ |
|---|---|---|---|---|---|---|
| $m_x'$ | (F,A,F)/ F/F/F//F | (F,S,F)/ F/F/F//F | (F,S,F)/ F/F/F//F | (F,A,F)/ F/F/F//F | (F,S,F)/F/ F/F//F | (F,S,F)/ F/F/F//F |
| $m_y'$ | (F,F,S)/ F/F/F//F | (F,F,A)/ F/F/F//F | (F,F,S)/ F/F/F//F | (F,F,S)/ F/F/F//F | (F,F,A)/ F/F/F//F | (F,F,S)/ F/F/F//F |
| $m_z'$ | (S,F,F)/ F/F/F//F | (S,F,F)/ F/F/F//F | (A,F,F)/ F/F/F//F | (S,F,F)/ F/F/F//F | (S,F,F)/F/ F/F//F | (A,F,F)/ F/F/F//F |
| $\bar{1}'$ | (F,F,F)/ F/F/F//A | (F,F,F)/ F/F/F//A | (F,F,F)/ F/F/F//A | (F,F,F)/ F/F/F//A | (F,F,F)/F/ F/F//A | (F,F,F)/ F/F/F//A |

The building of the symmetry classification of BL and BP is similar, and even the set of groups is the same. The only difference in the case of BP is that operations of the type $g^{(1)}$ do not prohibit the existence of any magnetization and polarization components, since the spatial inhomogeneities of the order parameters in the case of BP have become essentially three-dimensional.

There are 11 point groups of magnetic symmetry of BP with t-even chirality (Tables 2.9, 2.10). The opposite type of chirality corresponds to the same number of other magnetic point groups (Tables 2.11, 2.12).

**Table 2.12.** Point groups of BP magnetic symmetry with t-odd chirality. Groups of the fourth and eighth order.

| **Group** | $M_x$ | $M_y$ | $M_z$ | $P_x$ | $P_y$ | $P_z$ |
|---|---|---|---|---|---|---|
| $m_x^{'}m_z^{'}2_y$ | (S,A,F)/A/F/F//F | (S,S,F)/S/F/F//F | (A,S,F)/A/F/F//F | (S,A,F)/A/F/F//F | (S,S,F)/S/F/F//F | (A,S,F)/A/F/F//F |
| $m_y^{'}m_z^{'}2_x$ | (S,F,S)/F/F/S//F | (S,F,A)/F/F/A//F | (A,F,S)/F/F/A//F | (S,F,S)/F/F/S//F | (S,F,A)/F/F/A//F | (A,F,S)/F/F/A//F |
| $2_x/m_x{'}$ | (F,A,F)/F/F/S//A | (F,S,F)/F/F/A//A | (F,S,F)/F/F/A//A | (F,A,F)/F/F/S//A | (F,S,F)/F/F/A//A | (F,S,F)/F/F/A//A |
| $2_y/m_y{'}$ | (F,F,S)/A/F/F//A | (F,F,A)/S/F/F//A | (F,F,S)/A/F/F//A | (F,F,S)/A/F/F//A | (F,F,A)/S/F/F//A | (F,F,S)/A/F/F//A |
| $2_z/m_z^{'}$ | (S,F,F)/F/A/F//A | (S,F,F)/F/A/F//A | (A,F,F)/F/S/F//A | (S,F,F)/F/A/F//A | (S,F,F)/F/A/F//A | (A,F,F)/F/S/F//A |
| $m_x^{'}m_y^{'}2_z$ | (F,A,S)/F/A/F//F | (F,S,A)/F/A/F//F | (F,S,S)/F/S/F//F | (F,A,S)/F/A/F//F | (F,S,A)/F/A/F//F | (F,S,S)/F/S/F//F |
| $m_x^{'}m_y^{'}m_z^{'}$ | (S,A,S)/A/A/S//A | (S,S,A)/S/A/A//A | (A,S,S)/A/S/A//A | (S,A,S)/A/A/S//A | (S,S,A)/S/A/A//A | (A,S,S)/A/S/A//A |

In the case of achiral BPs, the same pattern is preserved. There are four point groups of magnetic symmetry of BPs of the second order (Table 2.13). The groups of the fourth order could be systematized by a crystal system: monoclinic (Table 2.15) and orthorhombic (Table 2.16), which correspond to nine and eight groups, respectively.

**Table 2.13.** Point groups of magnetic symmetry of achiral BPs. Groups of the second order.

| **Group** | $M_x$ | $M_y$ | $M_z$ | $P_x$ | $P_y$ | $P_z$ |
|---|---|---|---|---|---|---|
| $m_y$ | (F,F,A)/F/F/F//F | (F,F,S)/F/F/F//F | (F,F,A)/F/F/F//F | (F,F,S)/F/F/F//F | (F,F,A)/F/F/F//F | (F,F,S)/F/F/F//F |
| $m_x$ | (F,S,F)/F/F/F//F | (F,A,F)/F/F/F//F | (F,A,F)/F/F/F//F | (F,A,F)/F/F/F//F | (F,S,F)/F/F/F//F | (F,S,F)/F/F/F//F |
| $m_z$ | (A,F,F)/F/F/F//F | (A,F,F)/F/F/F//F | (S,F,F)/F/F/F//F | (S,F,F)/F/F/F//F | (S,F,F)/F/F/F//F | (A,F,F)/F/F/F//F |
| $\bar{1}$ | (F,F,F)/F/F/F//S | (F,F,F)/F/F/F//S | (F,F,F)/F/F/F//S | (F,F,F)/F/F/F//A | (F,F,F)/F/F/F//A | (F,F,F)/F/F/F//A |

The point symmetry groups of the second order of achiral BPs (Table 2.13) correspond to changes in one or all spatial coordinates. Accordingly, only one dependence of a spatial coordinate is determined.

**Table 2.14.** Point groups of magnetic symmetry of achiral BPs. Groups of the eighth order.

| **Group** | $M_x$ | $M_y$ | $M_z$ | $P_x$ | $P_y$ | $P_z$ |
|---|---|---|---|---|---|---|
| $m_x m_z m_y'$ | (A,S,S)/ A/S/A//A | (A,A,A)/ S/S/S//A | (S,A,S)/ A/A/S//A | (S,A,S)/ A/A/S//A | (S,S,A)/ S/A/A//A | (A,S,S)/ A/S/A//A |
| $m_y m_z m_x'$ | (A,A,A)/ S/S/S//A | (A,S,S)/ A/S/A//A | (S,S,A)/ S/A/A//A | (S,A,S)/ A/A/S//A | (S,S,A)/ S/A/A//A | (A,S,S)/ A/S/A//A |
| $m_y m_x' m_z'$ | (S,A,A)/ A/S/A//S | (S,S,S)/S/ S/S//S | (A,S,A)/ A/A/S//S | (S,A,S)/ A/A/S//A | (S,S,A)/ S/A/A//A | (A,S,S)/ A/S/A//A |
| $m_x m_y' m_z'$ | (S,S,S)/S/ S/S//S | (S,A,A)/ A/S/A//S | (A,A,S)/ S/A/A//S | (S,A,S)/ A/A/S//A | (S,S,A)/ S/A/A//A | (A,S,S)/ A/S/A//A |
| $m_z m_x' m_y'$ | (A,A,S)/ S/A/A//S | (A,S,A)/ A/A/S//S | (S,S,S)/S/ S/S//S | (S,A,S)/ A/A/S//A | (S,S,A)/ S/A/A//A | (A,S,S)/ A/S/A//A |

The point symmetry groups of the eighth order of achiral BPs (Table 2.14) correspond to the definition of all functional dependencies of spatial coordinates. The fourth-order groups of monoclinic and orthorhombic crystal systems always allow four dependencies of arbitrary type *F* (Tables 2.15-2.16).

**Table 2.15.** Point groups of magnetic symmetry of achiral BPs. Groups of the fourth order of the monoclinic crystal system.

| **Group** | $M_x$ | $M_y$ | $M_z$ | $P_x$ | $P_y$ | $P_z$ |
|---|---|---|---|---|---|---|
| $2'_x/m_x$ | (F,S,F)/F/F/A//A | (F,A,F)/F/F/S//A | (F,A,F)/F/F/S//A | (F,A,F)/F/F/S//A | (F,S,F)/F/F/A//A | (F,S,F)/F/F/A//A |
| $2'_y/m_y$ | (F,F,A)/S/F/F//A | (F,F,S)/A/F/F//A | (F,F,A)/S/F/F//A | (F,F,S)/A/F/F//A | (F,F,A)/S/F/F//A | (F,F,S)/A/F/F//A |
| $2'_z/m_z$ | (A,F,F)/F/S/F//A | (A,F,F)/F/S/F//A | (S,F,F)/F/A/F//A | (S,F,F)/F/A/F//A | (S,F,F)/F/A/F//A | (A,F,F)/F/S/F//A |
| $2_y/m_y$ | (F,F,A)/A/F/F//S | (F,F,S)/S/F/F//S | (F,F,A)/A/F/F//S | (F,F,S)/A/F/F//A | (F,F,A)/S/F/F//A | (F,F,S)/A/F/F//A |
| $2_x/m_x$ | (F,S,F)/F/F/S//S | (F,A,F)/F/F/A//S | (F,A,F)/F/F/A//S | (F,A,F)/F/F/S//A | (F,S,F)/F/F/A//A | (F,S,F)/F/F/A//A |
| $2'_z/m'_z$ | (S,F,F)/F/S/F//S | (S,F,F)/F/S/F//S | (A,F,F)/F/A/F//S | (S,F,F)/F/A/F//A | (S,F,F)/F/A/F//A | (A,F,F)/F/S/F//A |
| $2'_x/m'_x$ | (F,A,F)/F/F/A//S | (F,S,F)/F/F/S//S | (F,S,F)/F/F/S//S | (F,A,F)/F/F/S//A | (F,S,F)/F/F/A//A | (F,S,F)/F/F/A//A |
| $2'_y/m'_y$ | (F,F,S)/S/F/F//S | (F,F,A)/A/F/F//S | (F,F,S)/S/F/F//S | (F,F,S)/A/F/F//A | (F,F,A)/S/F/F//A | (F,F,S)/A/F/F//A |
| $2_z/m_z$ | (A,F,F)/F/A/F//S | (A,F,F)/F/A/F//S | (S,F,F)/F/S/F//S | (S,F,F)/F/A/F//A | (S,F,F)/F/A/F//A | (A,F,F)/F/S/F//A |

In the monoclinic crystal system, the coordinate dependence for BP volume is always determined $T_{zxy}$ (Table 2.15). For the point groups of magnetic symmetry of the orthorhombic crystal system, this dependence is always of type *F* (Table 2.16). In total, there are 26 groups of magnetic symmetry of achiral BPs (Tables 2.13-2.16).

**Table 2.16.** Point groups of magnetic symmetry of achiral BPs. Groups of the fourth order of the rhombic crystal system.

| **Group** | $M_x$ | $M_y$ | $M_z$ | $P_x$ | $P_y$ | $P_z$ |
|---|---|---|---|---|---|---|
| $m'_y m_x 2'_z$ | (F,S,S)/F/S/F//F | (F,A,A)/F/S/F//F | (F,A,S)/F/A/F//F | (F,A,S)/F/A/F//F | (F,S,A)/F/A/F//F | (F,S,S)/F/S/F//F |
| $m'_x m_y 2'_z$ | (F,A,A)/F/S/F//F | (F,S,S)/F/S/F//F | (F,S,A)/F/A/F//F | (F,A,S)/F/A/F//F | (F,S,A)/F/A/F//F | (F,S,S)/F/S/F//F |
| $m_x m_z 2_y$ | (A,S,F)/A/F/F//F | (A,A,F)/S/F/F//F | (S,A,F)/A/F/F//F | (S,A,F)/A/F/F//F | (S,S,F)/S/F/F//F | (A,S,F)/A/F/F//F |
| $m_y m_z 2_x$ | (A,F,A)/F/F/S//F | (A,F,S)/F/F/A//F | (S,F,A)/F/F/A//F | (S,F,S)/F/F/S//F | (S,F,A)/F/F/A//F | (A,F,S)/F/F/A//F |
| $m_z m'_y 2'_x$ | (A,F,S)/F/F/A//F | (A,F,A)/F/F/S//F | (S,F,S)/F/F/S//F | (S,F,S)/F/F/S//F | (S,F,A)/F/F/A//F | (A,F,S)/F/F/A//F |
| $m_z m'_x 2'_y$ | (A,A,F)/S/F/F//F | (A,S,F)/A/F/F//F | (S,S,F)/S/F/F//F | (S,A,F)/A/F/F//F | (S,S,F)/S/F/F//F | (A,S,F)/A/F/F//F |
| $m_y m'_z 2'_x$ | (S,F,A)/F/F/A//F | (S,F,S)/F/F/S//F | (A,F,A)/F/F/S//F | (S,F,S)/F/F/S//F | (S,F,A)/F/F/A//F | (A,F,S)/F/F/A//F |
| $m_x m'_z 2'_y$ | (S,S,F)/S/F/F//F | (S,A,F)/A/F/F//F | (A,A,F)/S/F/F//F | (S,A,F)/A/F/F//F | (S,S,F)/S/F/F//F | (A,S,F)/A/F/F//F |

## 2.4. Symmetry classification of the vortex magnetization distribution in the plates/films

To describe the vortex type of a magnetization distribution in the plates/films, we choose the following ansatz:

$$\boldsymbol{M}_{\perp}(\rho,\varphi)=M_s\left(\boldsymbol{e}_x\cos(n\varphi+c\pi/2)+\boldsymbol{e}_y\sin(n\varphi+c\pi/2)\right), \tag{2.29}$$

where $M_s$ - is the saturation magnetization and $n$ - is the vorticity. The parameter $c$ is defined within the limits:

$$-2\leq c\leq 2 \tag{2.30}$$

First, we should investigate the symmetry properties of the ansatz (2.29) itself (the point group of magnetic symmetry $G_V$), considering it as a spatial distribution of axial t-odd vectors.

The value of $n=1$ corresponds to a simple vortex state (Fig. 2.2 (a)-(c)). Such structures have symmetry elements $\infty_z$ and $m_z^{'}$which generate the boundary group given by magnetic type of the limit Curie groups. If $c=\pm1$ then the point groups contain mirror reflections $m_{\perp}$orthogonal to the plane $(XY)$ (Fig. 2.2 (c)). If $c=\pm 2$ or $c=0$ then some of the mirror reflections are supplemented by the time reversal operation ($m_{\perp}^{'}$) (Fig. 2.2 (a)). If the parameter $c$ is not an integer, then similar (the plane of the mirror contains $Z$) mirror reflections are absent in the group $G_V$ (Fig. 2.2 (b)).

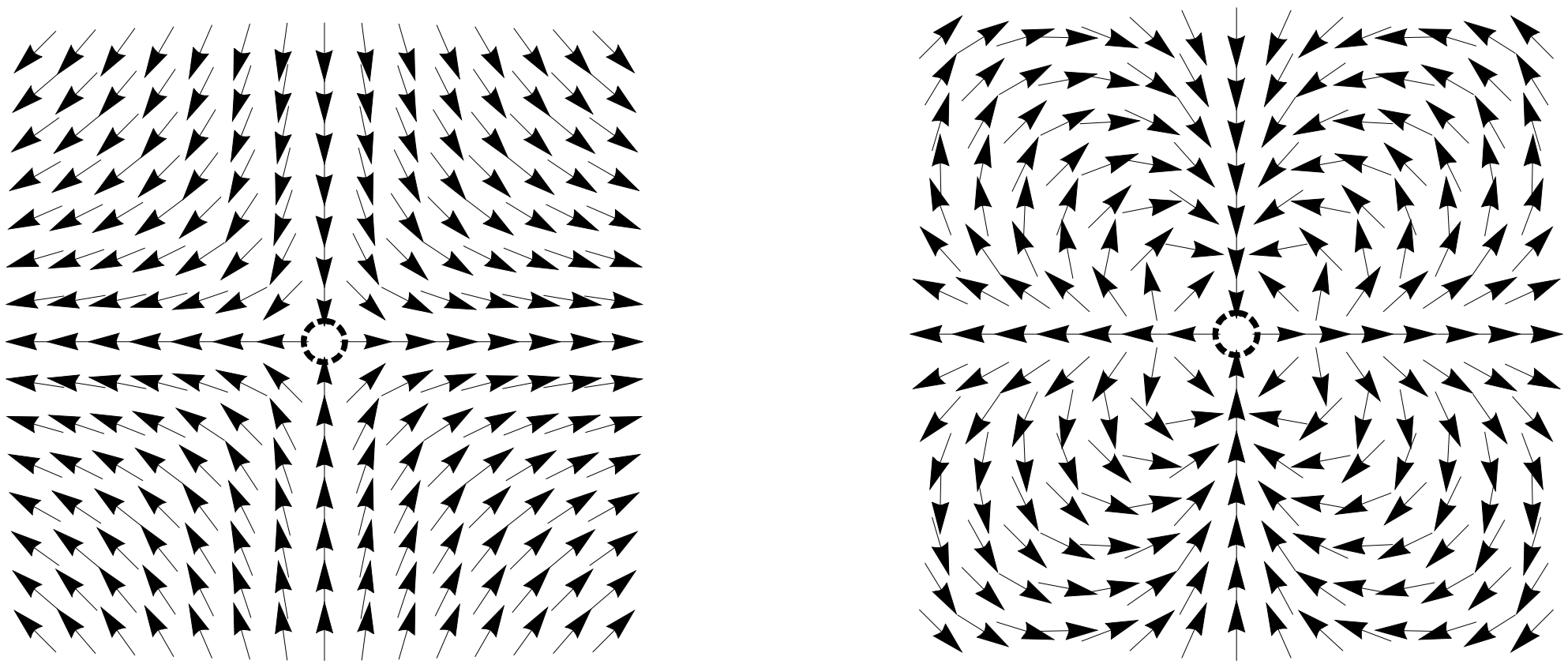

**(a)** **(b)**

**Fig. 2.1.** Schematic illustration of the magnetization distribution for the cases of the symmetry group $4'/m_z^{'} m_{\perp}^{'} m_{\perp}$ with the vorticity $n$ equal to -1 (a) and 3 (b).

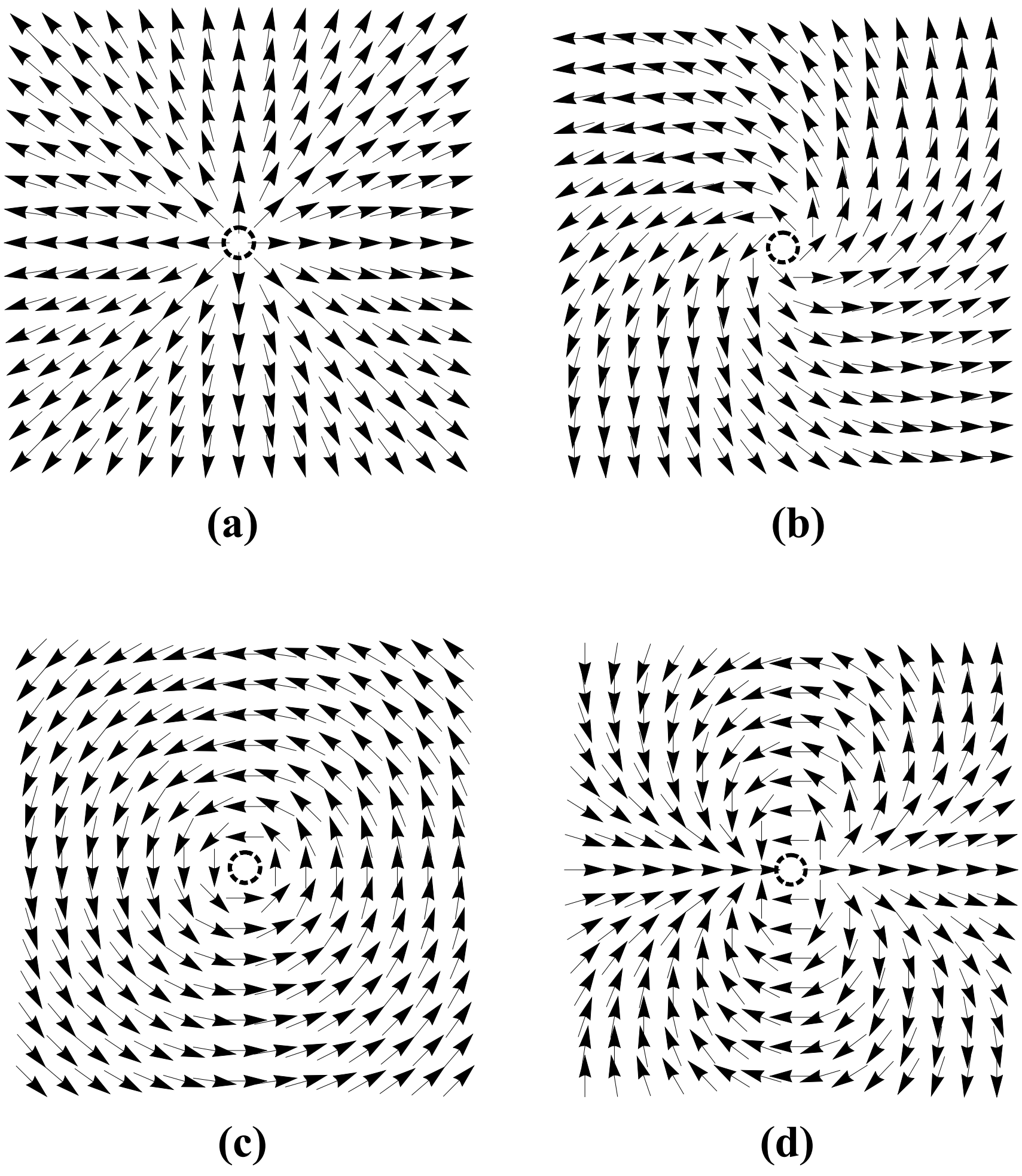

**Fig. 2.2.** Schematic illustration of the magnetization distribution for the cases of symmetry: $\infty_z / m_z' m_x' m_{xy}'$ (a), $\infty_z / m_z'$ (b), $\infty_z / m_z' m_x m_{xy}$ (c) and $m_y m_x' m_z'$ (d).

The following range:

$$n \in (-\infty, -1] \cup [2, +\infty) \qquad (2.31)$$

corresponds to structures without axial symmetry $\infty_z$ (Fig. 2.2 (d), Fig. 2.1). In this case, the group $G_V$ always contains a mirror image $m_z'$ and a rotation axis $r'$ where the order is as follows:

$$r = 2|n-1| \qquad (2.32)$$

The group $G_V$ contains $|n-1|$ mirrors $m_{\perp}$ and the same number of $m'_{\perp}$. Changes to the parameter $c$ at $n \neq 1$ lead to a rotation of the structure around the axis $Z$. Symmetry classification of groups $G_V$ was constructed using the listed symmetry operations (Table 2.17). The theory of magnetic limit symmetry [117] and Barron's chirality definition were used. Structures with t-even chirality have the corresponding degeneration: right- and left-helical states.

If there is a component of type:

$$M_z(\rho) \neq 0, \tag{2.33}$$

the group $G_V$ is transferred to the group $G_C$ which does not contain operations that change this component.

**Table 2.17.** Symmetry classification of vortex magnetization distribution

| $n$ | $c$ | $G_V$ | $G_C$ | **Chirality** |
|---|---|---|---|---|
| 1 | $\pm 1$ | $\infty_z / m_z' m_x m_{xy}$ | $\infty_z 2_x' 2_{xy}'$ | t-even |
| 1 | $0, \pm 2$ | $\infty_z / m_z' m_x' m_{xy}'$ | $\infty_z m_x' m_{xy}'$ | t-odd |
| 1 | $c \notin Z$ | $\infty_z / m_z'$ | $\infty_z$ | t-even |
| 2 | $\forall c$ | $m_y m_x' m_z'$ | $2_x' / m_x'$ | missing |
| all other values | $\forall c$ | $(2\|n-1\|)_z' / m_z' m_\xi' m_\eta$ *) | $\|n-1\|_z m_\xi' m_\eta'$ (odd $n$) | t-odd |
| | | | $\overline{\|n-1\|}_z m_\xi'$ (even $n$) | missing |

*) planes of the mirrors $m_\xi'$ and $m_\eta'$ perpendicular to the plane $(XY)$.

To obtain the symmetry groups $\widetilde{G}_V$ and $\widetilde{G}_C$ of real physical properties, it is necessary to intersect the corresponding groups $G_V$ and $G_C$ with the group $G_P$. In the case of constructing the group $G_C$, the symmetry elements $h_i$ that change the axial t-odd vector $M_z$ must be removed from the symmetry group. The set of these operations $\{h_i\}$ includes inversion at the center $\bar{1}'$, rotational axes $r'$ or inverse rotational axes $\bar{r}'$, mirroring $m_z'$ and mirror reflections, which include the axis $Z$: $m_x$, $m_y$, $m_{xy}$ and others. The rotational axis $|n-1|$-th order arises instead of $r$.

Structures with $n=-1$ arise in samples of the crosslike type [118,119]. The symmetry of the central part of such samples (magnetic shape symmetry) is as follows:

$$G_S = 4/mmm1', \tag{2.34}$$

which affects the nature of the magnetization distribution. Accordingly, the subgroup that allows this type of structure is:

$$4_z'/m_z' m_\xi' m_\eta \subseteq G_S, \tag{2.35}$$

which coincides with the group obtained (Table 2.17) for $n=-1$ based on the ansatz (2.29).

### 2.5. Conclusions

In this section, a symmetry classification of all main micromagnetic structures in magnetically ordered crystals is proposed. On the basis of the group-theoretical analysis, lists of all possible point groups of magnetic symmetry were obtained, and the spatial distribution of order parameters for each symmetry group was qualitatively characterized. The following conclusions have been made:

1. In total, there are 64 groups of magnetic symmetry of planar magnetic DWs. Among them, 42 groups correspond to 180° DWs, 42 groups – to 0° DWs, and 10 groups – to other types of DWs.

2. In total, there are 48 groups of magnetic symmetry of BLs, including 22 (11 t-even and 11 t-odd) chiral and 26 achiral ones.

3. The symmetry classification of BPs is similar to the case of BLs and consists of the same 48 classes.

4. There are 3 groups of symmetry of the vortex magnetization distribution in the sample plane with axial symmetry (rotation axis $\infty_z$). In addition to them, an infinite set of groups has the power of the set of natural numbers and tends to the limiting Curie groups of magnetic symmetry.

# SECTION 3

# SYMMETRY AND DEGENERATION OF MICROMAGNETIC STRUCTURES IN SPATIALLY CONFINED SAMPLES

## 3.1. Effect of the sample surface on the symmetry of the domain wall.

As noted in the previous section, the proposed symmetric analysis of micromagnetic structures requires adaptation to account for the influence of the crystal sample shape. This influence can be of a different nature, including any nonlocal interactions: shape anisotropy (demagnetization fields), magnetostriction (spatial stress distribution depends on the shape in the equilibrium state), surface exchange interaction [120], etc. The key channel for such an influence is also the boundary conditions of the micromagnetic variational problem: the idealization of an unbounded crystal in which DW, BL, and other micromagnetic structures are considered is unacceptably coarse in many cases. Above, we have considered MAVs, which turn out to be equilibrium in samples of limited size. Group-theoretical methods allow one to automatically account all the effects of sample surfaces at the level of qualitative (symmetric) analysis regardless of the effect's nature.

Such an analysis is reduced to finding the point group of magnetic symmetry $G_S$ and considering the Curie principle of symmetry of the paramagnetic phase (2.8). This approach can be used for arbitrary micromagnetic structures. Here, we show the relevant properties of the symmetry classification on the constructive example of magnetic DWs in (*nml*) plate/film of cubic $m\bar{3}m$ crystals.

A single coordinate system for all types of magnetic DWs can be defined through basis of the following form:

$$\left[\boldsymbol{e}_{\tilde{x}}, \boldsymbol{e}_{\tilde{y}}, \boldsymbol{e}_{\tilde{z}}\right] = \left[\boldsymbol{a}_2, -\boldsymbol{a}_1, \boldsymbol{n}_W\right] , \tag{3.1}$$

where $\boldsymbol{n}_W$ is the normal to the DW plane. For the 180° DW case, the vectors $\boldsymbol{a}_1$ and $\boldsymbol{a}_2$ are defined in [37] as $\boldsymbol{\tau}_1$ and $\boldsymbol{\tau}_2$, respectively. For the case of the angle $2\alpha \neq 180°$ between domains' magnetization directions $\boldsymbol{m}_1$ and $\boldsymbol{m}_2$, the unit vector $\boldsymbol{a}_1$ is selected as:

$$\boldsymbol{a}_1 = \left[\Delta\boldsymbol{m} - \boldsymbol{n}_W\left(\boldsymbol{n}_W \Delta\boldsymbol{m}\right)\right] / \left|\Delta\boldsymbol{m} - \boldsymbol{n}_W\left(\boldsymbol{n}_W \Delta\boldsymbol{m}\right)\right| \qquad (3.2)$$

at $b_\Delta \neq 0$ and $b_\Sigma = 0$, or by the ratio:

$$\boldsymbol{a}_1 = \left[\boldsymbol{a}_2 \times \boldsymbol{n}_W\right] \qquad (3.3)$$

at $b_\Delta = 0$ or $b_\Sigma \neq 0$, where the relevant parameters are set by ratios :

$$\Delta\boldsymbol{m} = \boldsymbol{m}_2 - \boldsymbol{m}_1 \qquad (3.4)$$

$$b_\Delta = \left|\left[\boldsymbol{n}_W \times \Delta\boldsymbol{m}\right]\right| \qquad (3.5)$$

$$b_\Sigma = \left|\left[\boldsymbol{n}_W \times \boldsymbol{m}_\Sigma\right]\right| \qquad (3.6)$$

The unit vector $\boldsymbol{a}_2$ is given by the expression:

$$\boldsymbol{a}_2 = \left[\boldsymbol{m}_\Sigma - \boldsymbol{n}_W\left(\boldsymbol{n}_W \boldsymbol{m}_\Sigma\right)\right] / \left|\boldsymbol{m}_\Sigma - \boldsymbol{n}_W\left(\boldsymbol{n}_W \boldsymbol{m}_\Sigma\right)\right| \qquad (3.7)$$

at $b_\Sigma \neq 0$ or by the following expression:

$$\boldsymbol{a}_2 = \left[\boldsymbol{n}_W \times \boldsymbol{a}_1\right] \qquad (3.8)$$

at $b_\Delta \neq 0$ and $b_\Sigma = 0$, or also with a direction in the DW plane ($a_2 \perp n_W$ at $b_\Sigma = b_\Delta = 0$), where the total vector is:

$$\boldsymbol{m}_\Sigma = \boldsymbol{m}_1 + \boldsymbol{m}_2 \qquad (3.9)$$

The relative orientation of the vectors $\boldsymbol{m}_1$, $\boldsymbol{m}_2$, and $\boldsymbol{n}_W$ is determined by the following parameters:

$$a_\Sigma=(\boldsymbol{n}_W \boldsymbol{m}_\Sigma) \quad (3.10)$$

$$a_\Delta=(\boldsymbol{n}_W \Delta \boldsymbol{m}) \quad (3.11)$$

$$a_C=(\boldsymbol{n}_W \boldsymbol{m}_C) \quad (3.12)$$

and the parameters $b_\Sigma$ and $b_\Delta$, where the vector is entered:

$$\boldsymbol{m}_C=[\boldsymbol{m}_1 \times \boldsymbol{m}_2] \quad (3.13)$$

The relative orientation of $\boldsymbol{m}_1$, $\boldsymbol{m}_2$, $\boldsymbol{n}_W$, and $\boldsymbol{n}_S$ is determined by the following parameters:

$$a_1=(\boldsymbol{a}_1 \boldsymbol{n}_S) \quad (3.14)$$

$$a_2=(\boldsymbol{a}_2 \boldsymbol{n}_S) \quad (3.15)$$

$$a_n=(\boldsymbol{n}_W \boldsymbol{n}_S) \quad (3.16)$$

$$b_1=|[\boldsymbol{n}_S \times \boldsymbol{a}_1]| \quad (3.17)$$

$$b_2=|[\boldsymbol{n}_S \times \boldsymbol{a}_2]| \quad (3.18)$$

$$b_n=|[\boldsymbol{n}_S \times \boldsymbol{n}_W]| \quad (3.19)$$

where $\boldsymbol{n}_S$ is the normal to the *(nml)* plane of the magnetic crystal sample.

The general list of 64 symmetry groups of magnetic DWs includes 42 groups ($1 \le k \le 42$ ) of 180° DWs, 10 groups ($7 \le k \le 13$ and $16 \le k \le 18$ ) of intermediate type DWs, and 42 groups ($k$ = 2, $6 \le k \le 13$, $16 \le k \le 19$, $k$ = 22, 24, 26, 30, 32, 37, 39, and$43 \le k \le 64$ ) of 0° DWs (numbering through $k$ is presented in Tables 3.1-3.4). Groups with symmetry axes of the 6th order are not considered, since they are absent in the $m\bar{3}m$ crystal.

Condition (2.8) allows us to find the relationship between the listed structural and orientation parameters and symmetry groups. All symmetry axes are parallel to the vectors $\boldsymbol{a}_1$ and $\boldsymbol{a}_2$. The mirroring planes are perpendicular to these axes. For the symmetry groups with numbers $k$ = 24, $k$ = 26, $k$ = 27, $29 \le k \le 36$, $42 \le k \le 51$, $k$ = 53, $55 \le k \le 60$ and $k$ = 64, only the symmetry elements that generate all other symmetry elements of the group are indicated.

**Table 3.1.** DW symmetry in spatially confined (111) samples.

| *k* | Sample orientation | Geometrical conditions | | Group |
|---|---|---|---|---|
| 24 | {111} | $b_n=0$ | $b_\Delta=b_\Sigma=0$ | $3_n$ |
| 26 | {111} | $b_n=0$ | $b_\Delta=b_\Sigma=0$ | $3_n, \bar{2}'_1$ |
| 27 | {111} | $b_n=0$ | $a_\Sigma=b_\Sigma=b_\Delta=0$ | $3_n, 2_1$ |
| 29 | {111} | $b_n=0$ | $a_\Sigma=b_\Sigma=b_\Delta=0$ | $\bar{3}'_n, \bar{2}'_1$ |
| 42 | {111} | $b_n=0$ | $a_\Sigma=b_\Sigma=b_\Delta=0$ | $\bar{3}'_n$ |
| 53 | {111} | $b_n=0$ | $a_\Delta=b_\Delta=b_\Sigma=0$ | $3_n, 2'_1$ |
| 55 | {111} | $b_n=0$ | $a_\Delta=b_\Delta=b_\Sigma=0$ | $\bar{3}_n, \bar{2}'_1$ |
| 64 | {111} | $b_n=0$ | $a_\Delta=b_\Delta=b_\Sigma=0$ | $\bar{3}_n$ |

**Table 3.2.** DW symmetry in spatially confined (100) and (110) samples.

| *k* | **Sample orientation** | **Geometrical conditions** | | **Group** |
|---|---|---|---|---|
| 1 | {100}, {110} | $b_n=0$ or $b_1=0$ | $a_\Sigma=b_\Sigma=a_\Delta=0$ | $(1,2_2,\bar{2}_1,\bar{2}_n)\times(1,\bar{1})$ |
| 2 | {100}, {110} | $b_n=0$ or $b_1=0$ | $a_\Delta=b_\Delta=a_\Sigma=0$, $a_\Sigma=b_\Sigma=a_\Delta=0$ | $1,\bar{2}'_1,\bar{2}_2,2'_n$ |
| 3 | {100}, {110} | $b_n=0$ or $b_1=0$ | $a_\Sigma=b_\Sigma=a_\Delta=0$ | $1,2_2,\bar{2}_1,\bar{2}_n$ |
| 7 | {100}, {110} | $b_n=0$ or $b_1=0$ | $a_\Sigma=a_\Delta=0$ | $1, 2'_1, 2_2, 2'_n$ |
| 9 | {100}, {110} | $b_n=0$ or $b_1=0$ | $a_C=a_\Delta=b_\Sigma=0$ | $1,2'_1,\bar{2}'_2,\bar{2}_n$ |
| 17 | {100}, {110} | $b_n=0$ or $b_1=0$ | $a_C=a_\Sigma=b_\Delta=0$ | $1,\bar{2}'_1,2_2,\bar{2}'_n$ |
| 21 | {100}, {110} | $b_n=0$ or $b_1=0$ | $a_\Sigma=b_\Sigma=b_\Delta=0$ | $1,2_1,2_2,2_n$ |
| 22 | {100}, {110} | $b_n=0$ or $b_1=0$ | $b_\Delta=b_\Sigma=0$ | $1,\bar{2}'_1,\bar{2}'_2,2_n$ |
| 23 | {100}, {110} | $b_n=0$ or $b_1=0$ | $a_\Sigma=b_\Sigma=b_\Delta=0$ | $(1,2_1,2_2,2_n)\times(1,\bar{1}')$ |
| 43 | {100}, {110} | $b_n=0$ or $b_1=0$ | $a_\Delta=b_\Delta=a_\Sigma=0$ | $(1,2'_1,\bar{2}_2,\bar{2}'_n)\times(1,\bar{1})$ |
| 44 | {100}, {110} | $b_n=0$ or $b_1=0$ | $a_\Delta=b_\Delta=a_\Sigma=0$ | $1,2'_1,\bar{2}_2,\bar{2}'_n$ |
| 50 | {100}, {110} | $b_n=0$ or $b_1=0$ | $a_\Delta=b_\Delta=b_\Sigma=0$ | $1,2'_1,2'_2,2_n$ |
| 51 | {100}, {110} | $b_n=0$ or $b_1=0$ | $a_\Delta=b_\Delta=b_\Sigma=0$ | $(1,2'_1,2'_2,2_n)\times(1,\bar{1})$ |

Separate consideration should be given to (*nml*)-films with arbitrary Miller indices except for non-zero values (Table 3.4):

$$|n| \neq |m| \neq |l| \neq |n| \tag{3.20}$$

The general list of 0° DW symmetry groups includes groups $k=2$, $6 \le k \le 13$, $16 \le k \le 19$, $k$=22, 24, 26, 30, 32, $43 \le k \le 51$, $k$=53,$55 \le k \le 60$ and $k$=64. The 60°- and 120° degree DWs are represented by groups $k$=10, 16, 18, and $k$=11, 13, 16, respectively.

**Table 3.3.** DW symmetry in spatially limited (100) samples.

| ***k*** | **Sample orientation** | **Geometrical conditions** | | **Group** |
|---|---|---|---|---|
| 30 | {100} | $b_n=0$ | $b_\Delta=b_\Sigma=0$ | $4_n$ |
| 31 | {100} | $b_n=0$ | $a_\Sigma=b_\Sigma=b_\Delta=0$ | $4_n,\bar{2}'_n$ |
| 32 | {100} | $b_n=0$ | $b_\Delta=b_\Sigma=0$ | $4_n,\bar{2}'_1$ |
| 33 | {100} | $b_n=0$ | $a_\Sigma=b_\Sigma=b_\Delta=0$ | $4_n,2_1$ |
| 34 | {100} | $b_n=0$ | $a_\Sigma=b_\Sigma=b_\Delta=0$ | $4_n,\bar{2}'_1,\bar{2}'_n$ |
| 35 | {100} | $b_n=0$ | $a_\Sigma=b_\Sigma=b_\Delta=0$ | $\bar{4}'_n$ |
| 36 | {100} | $b_n=0$ | $a_\Sigma=b_\Sigma=b_\Delta=0$ | $\bar{4}'_n,2_1$ |
| 56 | {100} | $b_n=0$ | $a_\Delta=b_\Delta=b_\Sigma=0$ | $4_n,\bar{2}_n$ |
| 57 | {100} | $b_n=0$ | $a_\Delta=b_\Delta=b_\Sigma=0$ | $4_n,2'_1$ |
| 58 | {100} | $b_n=0$ | $a_\Delta=b_\Delta=b_\Sigma=0$ | $4_n,\bar{2}'_1,\bar{2}_n$ |
| 59 | {100} | $b_n=0$ | $a_\Delta=b_\Delta=b_\Sigma=0$ | $\bar{4}_n$ |
| 60 | {100} | $b_n=0$ | $a_\Delta=b_\Delta=b_\Sigma=0$ | $\bar{4}_n,2'_1$ |

The symmetry groups 70.5°, 90° (for <100> or <110> axes of easy magnetization) and 109.5° DW correspond to the values $7<k<13$, $16<k<18$. The general list of 180° DW includes symmetry groups from $1 \le k \le 42$.

**Table 3.4.** DW symmetry in spatially limited (*nml)* samples.

| *k* | **Sample orientation** | **Geometrical conditions** | | **Group** |
|---|---|---|---|---|
| 4 | $\{nml\}$ | $a_1=0$ or $b_1=0$ | $a_\Sigma=b_\Sigma=a_\Delta=0$ | $1,\bar{1}',2'_1,\bar{2}_1$ |
| 5 | $\{nml\}$ | $a_n=0$ or $b_n=0$ | $a_\Sigma=b_\Sigma=a_\Delta=0$ | $1,\bar{1}',2'_n,\bar{2}_n$ |
| 6 | $\{nml\}$ | $a_2=0$ or $b_2=0$ | $a_\Sigma=a_\Delta=a_C=0$ | $1,\bar{2}_2$ |
| 8 | $\{nml\}$ | $a_n=0$ or $b_n=0$ | $a_\Sigma=a_\Delta=0$ | $1,2_n'$ |
| 10 | $\{nml\}$ | $a_1=0$ or $b_1=0$ | $a_\Delta=0$ | $1,2_1'$ |
| 11 | $\{nml\}$ | $a_n=0$ or $b_n=0$ | $a_C=a_\Delta=b_\Sigma=0$ | $1,\bar{2}_n$ |
| 12 | $\{nml\}$ | $a_1=0$ or $b_1=0$ | $a_C=0$ | $1,\bar{2}_1'$ |
| 13 | $\{nml\}$ | $a_2=0$ or $b_2=0$ | $a_\Sigma=0$ | $1,2_2$ |
| 14 | $\{nml\}$ | $a_2=0$ or $b_2=0$ | $a_\Sigma=b_\Sigma=0$ | $1,\bar{1},2_2',\bar{2}_2'$ |
| 15 | $\forall$ | $\forall$ | $a_\Sigma=b_\Sigma=0$ | $1,\bar{1}'$ |
| 16 | $\forall$ | $\forall$ | $\forall$ | 1 |
| 18 | $\{nml\}$ | $a_n=0$ or $b_n=0$ | $a_C=a_\Sigma=b_\Delta=0$ | $1,\bar{2}_n'$ |
| 19 | $\{nml\}$ | $a_n=0$ or $b_n=0$ | $b_\Delta=b_\Sigma=0$ | $1,2_n$ |
| 20 | $\{nml\}$ | $a_n=0$ or $b_n=0$ | $a_\Sigma=b_\Sigma=b_\Delta=0$ | $1,\bar{1}',2_n,\bar{2}_n'$ |
| 45 | $\{nml\}$ | $a_2=0$ or $b_2=0$ | $a_\Delta=b_\Delta=a_\Sigma=0$ | $1,\bar{1},2_2,\bar{2}_2$ |
| 46 | $\{nml\}$ | $a_n=0$ or $b_n=0$ | $a_\Delta=b_\Delta=a_\Sigma=0$ | $1,\bar{1},2_n',\bar{2}_n'$ |
| 47 | $\{nml\}$ | $a_1=0$ or $b_1=0$ | $a_\Delta=b_\Delta=0$ | $1,\bar{1},2_1',\bar{2}_1'$ |
| 48 | $\forall$ | $\forall$ | $a_\Delta=b_\Delta=0$ | $1,\bar{1}$ |
| 49 | $\{nml\}$ | $a_n=0$ or $b_n=0$ | $a_\Delta=b_\Delta=b_\Sigma=0$ | $1,\bar{1},2_n,\bar{2}_n$ |

## 3.2 Degeneration of the magnetic domain wall.

From the symmetry classification of magnetic DWs in cubic crystal samples of arbitrary crystallographic orientation, the systematization of their degeneracy follows. For any magnetic DW, the degeneration is obtained by the expansion of the form (2.12) for the point group of magnetic symmetry of the paramagnetic phase of the sample (2.8). The corresponding symmetry groups and degrees of degeneracy are listed for DWs of different types in a spatially unbounded cubic crystal (Tables 3.5-3.7).

**Table 3.5.** The sequence number $k$ and degeneracy $q_B$ of the boundary conditions $2\alpha$ DW ($2\alpha > 90°$) in a cubic $m\bar{3}m$ crystal for different directions of domain magnetization (low symmetry orientation).

| **DW plane** | **Boundary conditions** | | | | |
|---|---|---|---|---|---|
| | 180° DW [100], $[\bar{1}00]$ | 180° DW [110], $[\bar{1}\bar{1}0]$ | 180° DW [111], $[\bar{1}\bar{1}\bar{1}]$ | 120°DW [110], $[0\bar{1}1]$ | 109° DW [111],$[1\bar{1}\bar{1}]$ |
| (*hkk*) | 14 (24) | 15 (48) | 14 (24) | 16 (96) | 12 (48) |
| $(\bar{h}hl)$ | 15 (48) | 4 (24) | 15 (48) | 16 (96) | 16 (96) |
| $(\bar{h}kh)$ | 15 (48) | 15 (48) | 15 (48) | 13 (48) | 16 (96) |
| $(h\bar{k}k)$ | 14 (24) | 15 (48) | 15 (48) | 16 (96) | 10 (48) |
| (*hk0*),$(\bar{h}k0)$ | 14 (24) | 14 (24) | 15 (48) | 16 (96) | 16 (96) |
| (*h0l*),$(\bar{h}0l)$ | 14 (24) | 15 (48) | 15 (48) | 16 (96) | 16 (96) |
| (*0kl)*,$(0\bar{k}l)$ | 4 (24) | 15 (48) | 15 (48) | 16 (96) | 13 (48) |
| (*hkl)*,$(\bar{h}kl)$,$(h\bar{k}l)$ $(hk\bar{l})$ | 15 (48) | 15 (48) | 15 (48) | 16 (96) | 16 (96) |

The orientation of the magnetic DW perpendicular to the crystallographic directions of low symmetry (Table 3.5) corresponds to the highest values of DW degeneracy. On the contrary, directions with high symmetry, such as [100], [110], and [111], correspond to lower degeneracy values (Table 3.6).

**Table 3.6.** The sequence number $k$ and degeneracy $q_B$ of the boundary conditions $2\alpha$ DW ($2\alpha > 90°$) in a cubic $m\bar{3}m$ crystal for different directions of domain magnetization (high symmetry orientation).

| DW plane | Boundary conditions | | | | |
|---|---|---|---|---|---|
| | 180° DW [100], $[\bar{1}00]$ | 180° DW [110], $[\bar{1}\bar{1}0]$ | 180° DW [111], $[\bar{1}\bar{1}\bar{1}]$ | 120° DW [110], $[0\bar{1}1]$ | 109° DW [111], $[1\bar{1}\bar{1}]$ |
| (100) | 34 (6) | 14 (24) | 14 (24) | 16 (96) | 9 (24) |
| (010) | 1 (12) | 14 (24) | 14 (24) | 13 (48) | 13 (48) |
| (001) | 1 (12) | 1 (12) | 14 (24) | 16 (96) | 13 (48) |
| (111) | 14 (24) | 14 (24) | 29 (8) | 16 (96) | 12 (48) |
| $(\bar{1}11)$ | 14 (24) | 4 (24) | 14 (24) | 13 (48) | 12 (48) |
| $(1\bar{1}1)$ | 14 (24) | 4 (24) | 14 (24) | 16 (96) | 10 (48) |
| $(11\bar{1})$ | 14 (24) | 14 (24) | 14 (24) | 13 (48) | 10 (48) |
| (110) | 14 (24) | 23 (12) | 14 (24) | 16 (96) | 16 (96) |
| (101) | 14 (24) | 15 (48) | 14 (24) | 11 (48) | 16 (96) |
| (011) | 1 (12) | 15 (48) | 14 (24) | 16 (96) | 17 (24) |
| $(\bar{1}10)$ | 14 (24) | 1 (12) | 5 (24) | 16 (96) | 16 (96) |
| $(\bar{1}01)$ | 14 (24) | 15 (48) | 5 (24) | 13 (48) | 16 (96) |
| $(0\bar{1}1)$ | 1 (12) | 15 (48) | 5 (24) | 16 (96) | 7 (24) |
| (*hhl*) | 15 (48) | 14 (24) | 14 (24) | 16 (96) | 16 (96) |
| (*hkh*) | 15 (48) | 15 (48) | 14 (24) | 16 (96) | 16 (96) |

In general, 180° DWs are characterized by lower degeneracy values than planar magnetic DWs of other types (Table 3.7): 60° DW, 71° DW, 90° DW, 109° DW, and120° DW.

**Table 3.7.** The sequence number $k$ and degeneracy $q_B$ of the boundary conditions $2\alpha$ DW ( $2\alpha \le 90°$) in a cubic $m\bar{3}m$ crystal for different directions of domain magnetization.

| DW plane | Boundary conditions | | | |
|---|---|---|---|---|
| | 90° DW [100],$[0\bar{1}0]$ | 90° DW [110], $[1\bar{1}0]$ | 71° DW [111], $[\bar{1}11]$ | 60° DW [110], [011] |
| (100) | 12 (48) | 9 (24) | 17 (24) | 16 (96) |
| (010) | 12 (48) | 17 (24) | 10 (48) | 10 (48) |
| (001) | 7 (24) | 7 (24) | 10 (48) | 16 (96) |
| (111) | 13 (48) | 16 (96) | 12 (48) | 10 (48) |
| $(\bar{1}11)$ | 10 (48) | 16 (96) | 12 (48) | 16 (96) |
| $(1\bar{1}1)$ | 10 (48) | 16 (96) | 13 (48) | 10 (48) |
| $(11\bar{1})$ | 13 (48) | 16 (96) | 13 (48) | 16 (96) |
| (110) | 17 (24) | 12 (48) | 16 (96) | 16 (96) |
| (101) | 16 (96) | 10 (48) | 16 (96) | 10 (48) |
| (011) | 16 (96) | 13 (48) | 9 (24) | 16 (96) |
| $(\bar{1}10)$ | 9 (24) | 12 (48) | 16 (96) | 16 (96) |
| $(\bar{1}01)$ | 16 (96) | 10 (48) | 16 (96) | 18 (48) |
| $(0\bar{1}1)$ | 16 (96) | 13 (48) | 7 (24) | 16 (96) |
| (*hhl*) | 13 (48) | 16 (96) | 16 (96) | 16 (96) |
| (*hkh*) | 16 (96) | 16 (96) | 16 (96) | 10 (48) |
| (*hkk*) | 16 (96) | 16 (96) | 12 (48) | 16 (96) |
| $(\bar{h}hl)$ | 10 (48) | 16 (96) | 16 (96) | 16 (96) |
| $(\bar{h}kh)$ | 16 (96) | 16 (96) | 16 (96) | 16 (96) |
| $(h\bar{k}k)$ | 16 (96) | 16 (96) | 13 (48) | 16 (96) |
| (*hk0*),$(\bar{h}k0)$ | 12 (48) | 12 (48) | 16 (96) | 16 (96) |
| (*h0l*),$(\bar{h}0l)$ | 16 (96) | 10 (48) | 16 (96) | 16 (96) |
| (*0kl*),$(0\bar{k}l)$ | 16 (96) | 13 (48) | 10 (48) | 16 (96) |
| (*hkl*),$(\bar{h}kl)$,$(h\bar{k}l)$, $(hk\bar{l})$ | 16 (96) | 16 (96) | 16 (96) | 16 (96) |

## 3.3 Degeneration of magnetic domain walls in a spatially confined crystal with negative magnetic anisotropy.

Let us consider the application of the constructed formal approach to specific DWs in a cubic crystal with negative magnetic anisotropy. It is necessary to consider arbitrary boundary conditions of 71°, 109°, and 180° DWs (hereinafter, $2\alpha$ DWs) in a plate-shaped crystal sample with an arbitrary crystallographic orientation. The type of DW is determined by the magnetization directions of neighboring domains:

$$2\alpha = \arccos\left(\boldsymbol{m}_1 \boldsymbol{m}_2\right) \tag{3.21}$$

The equilibrium orientations of planar magnetic DWs are determined by the minimum of their total free energy in the crystal, where the energy is derived from the energy of the corresponding volume of material with homogeneous magnetization in the direction of the axis of easy magnetization (domains). The total free energy is determined by the product of the surface energy density of the DWs and their area, which depends on the orientation of the DW plane relative to the plane of the free surfaces of the crystal sample. This is one of the channels of influence of the symmetry of the sample shape on the symmetry of the micromagnetic structure, which was formally investigated above. This channel exists even when there is no other surface effect, for example, in the case of flat Bloch DWs with neglected demagnetization field effect. Thus, the symmetry and degeneration of DWs in unconfined and confined crystals are different for most DW orientations.

Different boundary conditions of the DW can correspond to the same dependence of the DW area on its orientation relative to the axes of easy magnetization. These are the boundary conditions that correspond to symmetrically equally oriented magnetizations of domains relative to the sample surface. This allows us to divide the types of $2\alpha$ DW into subtypes of equal energy for a given sample, namely $Q_{2\alpha}$ subtypes. Similar domain walls from different subtypes have different equilibrium orientations, areas, and free energies.

Each subtype includes $q$ DWs of the same type with the same energies and equilibrium areas and different boundary conditions, i.e. different couples of vectors $\boldsymbol{m}_\Sigma$ and $\Delta\boldsymbol{m}$ which correspond to the same angles relative to the normal to the sample plane, respectively, the angles $\beta$ and $\gamma$:

$$\beta = \arccos\left|\boldsymbol{m}_\Sigma \boldsymbol{n}_S / (2\cos\alpha)\right| \tag{3.22}$$

$$\gamma = \arccos\left|\Delta\boldsymbol{m}\,\boldsymbol{n}_S / (2\sin\alpha)\right| \tag{3.23}$$

In general, the values of $q$ are different for different subtypes of DWs. In an arbitrarily oriented (*nml*) plate, there are two $71^0$ DWs (or two $109^0$ DWs) belonging to the same subtype if their vectors $\boldsymbol{m}_\Sigma$ (or $\Delta\boldsymbol{m}$, respectively) satisfy the conditions:

$$\boldsymbol{m}_\Sigma(1)\|\left(\boldsymbol{e}_i+\boldsymbol{e}_j\right)/\sqrt{2} \tag{3.24}$$

$$\boldsymbol{m}_\Sigma(2)\|\left(\boldsymbol{e}_i-\boldsymbol{e}_j\right)/\sqrt{2} \tag{3.25}$$

for cases (*nml*) plates/films that meet the conditions:

$$S_i S_j=0, \tag{3.25}$$

where

$$S_i=\left(\boldsymbol{n}_S \boldsymbol{e}_i\right) \tag{3.26}$$

$$S_j=\left(\boldsymbol{n}_S \boldsymbol{e}_j\right), \tag{3.27}$$

for $i\neq j, i,\ j=1,\ 2,\ 3$; or such conditions:

$$\boldsymbol{m}_\Sigma(1)\|\left(\boldsymbol{e}_i\pm\boldsymbol{e}_j\right)/\sqrt{2} \tag{3.28}$$

$$\boldsymbol{m}_\Sigma(2)\|\left(\boldsymbol{e}_i\pm\boldsymbol{e}_k\right)/\sqrt{2} \tag{3.29}$$

for cases (*nml*) of plates/films that satisfy either condition:

$$S_i=\pm S_j \tag{3.30}$$

$$S_i=\mp S_j=\mp S_k \tag{3.31}$$

$$\begin{cases} S_i=0 \\ S_j=-S_k \end{cases} \tag{3.32}$$

where $k\neq i,\ j,\ k=1,2$, and 3.

In the case of $180^0$ DW, domain walls belong to the same subtype if the following conditions are met:

$$\Delta\boldsymbol{m}(1)\|\left(\boldsymbol{e_i}\pm\boldsymbol{e_j}+\boldsymbol{e_k}\right)/\sqrt{3} \quad (3.33)$$

$$\Delta\boldsymbol{m}(2)\|\left(-\boldsymbol{e_i}\pm\boldsymbol{e_j}+\boldsymbol{e_k}\right)/\sqrt{3} \quad (3.34)$$

for cases (*nml*) plates/films that satisfy either condition:

$$S_i=0 \quad (3.35)$$

$$S_j=-S_k; \quad (3.36)$$

or, either, such conditions:

$$\Delta\boldsymbol{m}(1)\|\left(\boldsymbol{e_i}+\boldsymbol{e_j}-\boldsymbol{e_k}\right)/\sqrt{3} \quad (3.37)$$

$$\Delta\boldsymbol{m}(2)\|\left(-\boldsymbol{e_i}+\boldsymbol{e_j}-\boldsymbol{e_k}\right)/\sqrt{3} \quad (3.38)$$

for cases (*nml*) plates/films that satisfy either condition:

$$S_i=0 \quad (3.39)$$

$$S_j=S_k; \quad (3.40)$$

or, either, such conditions:

$$\Delta\boldsymbol{m}(1)\|\left(\boldsymbol{e_i}-\boldsymbol{e_j}-\boldsymbol{e_k}\right)/\sqrt{3} \quad (3.41)$$

$$\Delta\boldsymbol{m}(2)\|-\left(\boldsymbol{e_i}+\boldsymbol{e_j}+\boldsymbol{e_k}\right)/\sqrt{3} \quad (3.42)$$

for cases (*nml*) plates/films that satisfy either condition:

$$S_i=0 \quad (3.43)$$

$$S_j=-S_k \quad (3.44)$$

**Table 3.8.** Degeneracy of 180° DW in (*nml*) cubic crystal plates/films with negative magnetocrystalline anisotropy

<table>
<tr><th rowspan="3">$\pm\Delta\boldsymbol{m}$</th><th colspan="12">$n\geq0$, $m\geq0$, $l\geq0$</th></tr>
<tr><th colspan="2">$n=m=0$; $l\neq0$</th><th colspan="2">$n=m==l$</th><th colspan="2">$m=l\neq n=0$</th><th colspan="2">$m\neq l$; $m,l\neq0$; $n=0$</th><th colspan="2">$n=m\neq l\neq0$; $n\neq0$</th><th colspan="2">$n\neq m\neq l$; $n\neq l$; $n,m,l\neq0$</th></tr>
<tr><th>$\gamma$,°</th><th>$q$</th><th>$\gamma$,°</th><th>$q$</th><th>$\gamma$,°</th><th>$q$</th><th>$\gamma$</th><th>$q$</th><th>$\gamma$</th><th>$q$</th><th>$\gamma$</th><th>$q$</th></tr>
<tr><td>$[111]$</td><td rowspan="4">54.74</td><td rowspan="4">4</td><td>0</td><td>1</td><td rowspan="2">35.26</td><td rowspan="2">2</td><td rowspan="2">$\arccos\left[\frac{\lvert m+l\rvert}{\sqrt{3u}}\right]$</td><td rowspan="2">2</td><td>$\arccos\left[\frac{\lvert l+2n\rvert}{\sqrt{3u}}\right]$</td><td>1</td><td>$\arccos\left[\frac{\lvert n+m+l\rvert}{\sqrt{3u}}\right]$</td><td>1</td></tr>
<tr><td>$[\bar{1}11]$</td><td rowspan="3">71</td><td rowspan="3">3</td><td rowspan="2">$\arccos\left[\frac{\lvert l\rvert}{\sqrt{3u}}\right]$</td><td rowspan="2">2</td><td>$\arccos\left[\frac{\lvert m+l-n\rvert}{\sqrt{3u}}\right]$</td><td>1</td></tr>
<tr><td>$[1\bar{1}1]$</td><td rowspan="2">90</td><td rowspan="2">2</td><td rowspan="2">$\arccos\left[\frac{\lvert m-l\rvert}{\sqrt{3u}}\right]$</td><td rowspan="2">2</td><td>$\arccos\left[\frac{\lvert n-m+l\rvert}{\sqrt{3u}}\right]$</td><td>1</td></tr>
<tr><td>$[11\bar{1}]$</td><td>$\arccos\left[\frac{\lvert l-2n\rvert}{\sqrt{3u}}\right]$</td><td>1</td><td>$\arccos\left[\frac{\lvert n+m-l\rvert}{\sqrt{3u}}\right]$</td><td>1</td></tr>
<tr><td>$Q_{180}$</td><td colspan="2">**1**</td><td colspan="2">**2**</td><td colspan="2">**2**</td><td colspan="2">**2**</td><td colspan="2">**3**</td><td colspan="2">**4**</td></tr>
</table>

Here, $u=n^2+m^2+l^2$.

All possible crystal plate/film orientations are described by the following combinations of Miller indices: (100), (110), (111), (*nnl*), (nm0), and (*nml)*, where *n*, *m,* and *l* are non-zero values of different moduli. The results of applying the criteria presented above to all possible DWs of a cubic crystal with negative magnetocrystalline anisotropy are summarized in the tables (Tables 3.8-3.10).

**Table 3.9.** Degeneracy of 71° and 109° DW in (*nml*) cubic crystal plates/films with negative magnetocrystalline anisotropy containing <100> directions in the plate/film plane

| $\pm\boldsymbol{m}_{\Sigma}$ $(\pm\Delta\boldsymbol{m})$ for $2\alpha=71°$ or $\pm\Delta\boldsymbol{m}$ $(\pm\boldsymbol{m}_{\Sigma})$ for $2\alpha=109°$ | *n*≥0, *m*≥0, *l*≥0 | | | | | |
|---|---|---|---|---|---|---|
| | $n=m=0;l\neq0$ | | $n=m\neq l=0$ | | $n\neq m;\ n,m\neq0;l=0$ | |
| | $\beta\{\gamma\}$,° for $2\alpha=71°$ or $\{\gamma\beta\}$,° for $2\alpha=109°$ | $q$ | $\beta\{\gamma\}$,° for $2\alpha=71°$ or $\{\gamma\beta\}$,° for $2\alpha=109°$ | $q$ | $\beta\{\gamma\}$ for $2\alpha=71°$ or $\{\gamma\beta\}$ for $2\alpha=109°$ | $q$ |
| $[110]([001])$ | $90\{0\}$ | 4 | $0\{90\}$ | 2 | $\arccos\left[\frac{\lvert n+m\rvert}{\sqrt{2u}}\right]$ $\{\pi/2\}$ | 2 |
| $[1\bar{1}0]([001])$ | | | $90\{90\}$ | 2 | $\arccos\left[\frac{\lvert n-m\rvert}{\sqrt{2u}}\right]$ $\{\pi/2\}$ | 2 |
| $[011]([100])$ | $45\{90\}$ | 8 | $60\{45\}$ | 8 | $\arccos\left[\lvert m\rvert/\sqrt{2u}\right]$ $\{\arccos\left[\lvert n\rvert/\sqrt{u}\right]\}$ | 4 |
| $[01\bar{1}]([100])$ | | | | | | |
| $[101]([010])$ | | | | | $\arccos\left[\lvert n\rvert/\sqrt{2u}\right]$ $\{\arccos\left[\lvert m\rvert/\sqrt{u}\right]\}$ | 4 |
| $[\bar{1}01]([010])$ | | | | | | |
| $Q_{71}$ or $Q_{109}$ | **2** | | **3** | | **4** | |

The number of $Q_{2\alpha}$ subtypes in an (*nml*) plate/film increases as the normal $\boldsymbol{n}_S$ of the plate/film plane deviates from the highly symmetrical crystallographic directions <001>, <110>, and <111>. The minimum number of subtypes occurs in the case of (001) plate.

In spatially confined crystals, 71° DW corresponds to a certain 109° DW by law:

$$\begin{cases} \beta' = \gamma'' \\ \gamma' = \beta'' \end{cases} \quad (3.45)$$

where the angles with one and two strokes correspond to 71° and 109° DW, respectively.

**Table 3.10.** Degeneracy of 71° and 109° DW in *(nml)* cubic crystal plates with negative magnetocrystalline anisotropy, which do not contain directions of type <100> in the plate/film plane

<table>
<tr><td rowspan="3">$\pm\boldsymbol{m}_\Sigma$ ($\pm \Delta\boldsymbol{m}$) for $2\alpha=71°$ or $\pm\Delta\boldsymbol{m}$ ($\pm\boldsymbol{m}_\Sigma$) for $2\alpha=109°$</td><td colspan="6">$n\geq 0, m\geq 0, l\geq 0$</td></tr>
<tr><td colspan="2">$n=m=l$</td><td colspan="2">$n=m\neq l\neq 0;\ n\neq 0$</td><td colspan="2">$n\neq m\neq l; n\neq l; n,m,l\neq 0$</td></tr>
<tr><td>$\beta\{\gamma\}$,° for $2\alpha$=71° or $\{\gamma\beta\}$,° for $2\alpha$=109°</td><td>$q$</td><td>$\beta\{\gamma\}$ for $2\alpha$=71° or $\{\gamma\beta\}$ for $2\alpha$=109°</td><td>$q$</td><td>$\beta\{\gamma\}$ for $2\alpha$=71° or $\{\gamma\beta\}$ for $2\alpha$=109°</td><td>$q$</td></tr>
<tr><td>$[110]([001])$</td><td rowspan="3">35.26 {54.74}</td><td rowspan="3">6</td><td>$\arccos[|n|\sqrt{2}/\sqrt{u}]$ $\{\arccos[|l|/\sqrt{u}]\}$</td><td>2</td><td>$\arccos[|n+m|/\sqrt{2u}]$ $\{\arccos[|l|/\sqrt{u}]\}$</td><td>2</td></tr>
<tr><td>$[101]([010])$</td><td rowspan="2">$\arccos[|n+l|/\sqrt{2u}]$ $\{\arccos[|n|/\sqrt{u}]\}$</td><td rowspan="2">4</td><td>$\arccos[|n+l|/\sqrt{2u}]$ $\{\arccos[|m|/\sqrt{u}]\}$</td><td>2</td></tr>
<tr><td>$[011]([100])$</td><td>$\arccos[|m+l|/\sqrt{2u}]$ $\{\arccos[|n|/\sqrt{u}]\}$</td><td>2</td></tr>
<tr><td>$[\bar{1}10]([001])$</td><td rowspan="3">90 {54.74}</td><td rowspan="3">6</td><td>$\pi/2$ $\{\arccos[|l|/\sqrt{u}]\}$</td><td>2</td><td>$\arccos[|n-m|/\sqrt{2u}]$ $\{\arccos[|l|/\sqrt{u}]\}$</td><td>2</td></tr>
<tr><td>$[0\bar{1}1]([100])$</td><td rowspan="2">$\arccos[|n-l|/\sqrt{2u}]$ $\{\arccos[|n|/\sqrt{u}]\}$</td><td rowspan="2">4</td><td>$\arccos[|m-l|/\sqrt{2u}]$ $\{\arccos[|n|/\sqrt{u}]\}$</td><td>2</td></tr>
<tr><td>$[10\bar{1}]([010])$</td><td>$\arccos[|n-l|/\sqrt{2u}]$ $\{\arccos[|m|/\sqrt{u}]\}$</td><td>2</td></tr>
<tr><td>$Q_{71}$ or $Q_{109}$</td><td colspan="2">**2**</td><td colspan="2">**4**</td><td colspan="2">**6**</td></tr>
</table>

The vector $\boldsymbol{m}_C$ of one 71° DW is always parallel to the $\boldsymbol{m}_\Sigma$ of the other 71° DW in the (*nml*) plate/film. These 71° DWs have collinear vectors $\Delta\boldsymbol{m}$ and same angles $\gamma$ **.** Let us introduce the angle $\varphi$:

$$\varphi = \arccos|\boldsymbol{n}_S \boldsymbol{m}_C / \sin 2\alpha| \tag{3.46}$$

The suggested parameters have a constant relationship (only 2 parameters are independent, it is the pair of parameters that determines the DW subtype):

$$\cos 2\beta + \cos 2\gamma + \cos 2\varphi = -1 \tag{3.47}$$

There exists a pair of identical non-180° DWs with identical angle $\varphi$ (and different $\beta$ and/or $\gamma$ ) in (*nml*) plates/films that do not contain directions of type <110> and do contain directions of type <112> in their plane. These include (*nm0*) and (*nml*) plates that satisfy the conditions:

$$\mathrm{D} = (|n|-|m|)(|m|-|l|)(|l|-|n|) \neq 0 \tag{3.48}$$

$$T = (2|m|-|l \pm n|)(2|n|-|m \pm l|)(2|l|-|n \pm m|) = 0 \tag{3.49}$$

There are DWs with opposite directions of magnetization rotation (t-even chirality in the case of a planar Bloch DW) at the same arrangement of domains but different energies (chirality degeneracy is removed) in samples that satisfy the condition:

$$\mathrm{D} = 0 \tag{3.50}$$

These include samples such as (100), (110), (111), and (*nnl*) plates. For the 109° DW case, the set of such plates is wider. It also includes wafers with directions <100> in their plane (*nml=0*).

When the vector vector $\boldsymbol{m}_\Sigma$ of one 71° DW coincides with the vector $\boldsymbol{m}_C$ of another 71° DW, these DWs can have the same energies at opposite directions of magnetization rotation only simultaneously. The equilibrium 71° DW and 109°

DW have magnetization rotation in the planes of these DWs in the samples with $D$=0. Equilibrium 180° DWs with different orientations of their planes occur in samples with directions of type <110> and <112> in their plane:

$$\begin{cases} TD=0 \\ T\neq D \end{cases} \tag{3.51}$$

if there are directions of type <111> , which are deviated by 40° from the sample surface.

### 3.4. Conclusions

In this section, we suggested group-theoretical methods for accounting of influence of any nature on micromagnetic structures from the free surfaces of a sample. This makes it possible to determine qualitatively relevant changes in their structure and symmetry. Using the constructive example of magnetic DWs, we show the organic connection between the symmetry classification and degeneration of the micromagnetic structure in spatially confined samples of arbitrary crystallographic orientation. The following conclusions are made for magnetic DWs:

1. The degeneracy of magnetic DWs in plate/film samples of limited size is always even and lies between 2 and 96. The switching of the corresponding states corresponds to potential applications in spintronic devices.

2. Any symmetry group of magnetic DW can emerge at a particular crystallographic orientation *(nml)* of the sample.

3. The greatest variety of DW symmetry groups is observed in (100), (110), and (111) samples.

# SECTION 4

# INHOMOGENEOUS MAGNETOELECTRIC PROPERTIES OF MICROMAGNETIC STRUCTURES

## 4.1. Phenomenological theory of the inhomogeneous magnetoelectric effect in crystals of cubic, tetragonal and orthorhombic crystal systems.

It is necessary to distinguish between two independent causes of the Lifshitz invariants in (1.3). Typically, they are obtained after the extraction of full derivatives. In addition, they may be required by the symmetry of the crystal. The full derivatives of the form (1.4) and (1.6) are special cases of inhomogeneous (aka *flexomagnetoelectric*) ME interaction, so they cannot be generally removed from the variational problem that considers the unknown functions $\boldsymbol{M}(r)$ and $\boldsymbol{P}(r)$. Generalization of the scalar form of electric polarization induction to the tensor form is necessary in uniaxial and biaxial crystals:

$$\boldsymbol{P}_{ME} = -\hat{\chi}_e \left( \partial F_{ME} / \partial \boldsymbol{P} \right) \tag{4.1}$$

The structure of the free energy density expression (1.3) was obtained for cubic, tetragonal, and orthorhombic ferromagnetic and ferrimagnetic crystals (4.2-4.6). The method of such derivation is the selection of invariants using the representations of the crystallographic group of magnetic symmetry of the crystal in the paramagnetic phase. If one takes into account that magnetization is an axial t-odd vector, electric polarization is a polar t-odd vector, the operator $\nabla_k$ is analogous to multiplication by the $k$-th component of the polar t-even vector, then one can find all possible invariants of the form (1.3), which are allowed by the magnetic symmetry $G_P$ of the paramagnetic phase of the crystal (for example, the group $m\bar{3}m1'$ groups in the case of a cubic $m\bar{3}m$ crystal). Each term in Eq. $F_{ME}$ (1.3) must remain constant or change into another term of this expression under the action of any symmetry operation of the group $G_P$. Otherwise, the corresponding term $P_i M_j \nabla_k M_l$ must have a coefficient $\gamma_{ijkl} \equiv 0$.

For the case of symmetric crystallographic groups ($m\bar{3}m$, 432, $\bar{4}3m$) of cubic symmetry, the expression for the free energy density of inhomogeneous ME interactions is as follows:

$$F_{ME}^{S} = \tilde{\gamma}_1 P_i \nabla_i M_i^2 + \tilde{\gamma}_2 (\boldsymbol{P}\nabla) \boldsymbol{M}^2 + \tilde{\gamma}_3 \boldsymbol{P} (\boldsymbol{M}\nabla) \boldsymbol{M} + \tilde{\gamma}_4 (\boldsymbol{P}\boldsymbol{M})(\nabla \boldsymbol{M}) \qquad (4.2)$$

For the alternating groups ($m\bar{3}$, 23) of the cubic crystal system, the corresponding expression is as follows:

$$F_{ME}^{A} = \tilde{\gamma}_1 P_i \nabla_i M_i^2 + \left(\tilde{\gamma}_2' + \tilde{\gamma}_2'' \varepsilon_{ijk}\right) P_j \nabla_j M_k^2 +$$

$$+ \left(\tilde{\gamma}_3' + \tilde{\gamma}_3'' \varepsilon_{ijk}\right) P_j M_k \nabla_k M_j + \left(\tilde{\gamma}_4' + \tilde{\gamma}_4'' \varepsilon_{ijk}\right) P_j M_j \nabla_k M_k \qquad (4.3)$$

For dihedral groups ($4_z/m_z m_x m_{xy}$, $\bar{4}_z 2_x m_{xy}$, $4_z m_x m_y$, $4_z 2_x 2_y$) of the tetragonal crystal system, the free energy of inhomogeneous ME interactions is described by the expression:

$$F_{ME}^{D} = \left(\tilde{\gamma}_1' + \tilde{\gamma}_1'' \delta_{iz}\right) P_i \nabla_i M_i^2 + \left(\tilde{\gamma}_2' + \tilde{\gamma}_2'' \delta_{iz} + \tilde{\gamma}_2''' \delta_{jz}\right) P_i \nabla_i M_j^2 +$$

$$+ \left(\tilde{\gamma}_3' + \tilde{\gamma}_3'' \delta_{iz} + \tilde{\gamma}_3''' \delta_{jz}\right) P_i M_j \nabla_j M_i + \left(\tilde{\gamma}_4' + \tilde{\gamma}_4'' \delta_{iz} + \tilde{\gamma}_4''' \delta_{jz}\right) P_i M_i \nabla_j M_j \qquad (4.4)$$

For the case of cyclic groups ($4_z/m_z$, $\bar{4}_z$, $4_z$) of the tetragonal crystal system, the corresponding expression is as follows:

$$F_{ME}^{C} = F_{ME}^{D} +$$

$$+ \left(1 - \delta_{iz} - \delta_{jz}\right) \varepsilon_{ijk} \times$$

$$\times \left(\tilde{\gamma}_5 P_i M_i \nabla_i M_j + \tilde{\gamma}_6 P_i \nabla_j M_i^2 + \tilde{\gamma}_7 P_i M_j \nabla_i M_i + \tilde{\gamma}_8 P_j \nabla_i M_i^2\right) +$$

$$+ \tilde{\gamma}_9 P_z M_z \left(\nabla_x M_y - \nabla_y M_x\right) + \tilde{\gamma}_{10} P_z \left(M_x \nabla_z M_y - M_y \nabla_z M_x\right) +$$

$$\tilde{\gamma}_{11} P_z \left(M_x \nabla_y - M_y \nabla_x\right) M_z + \tilde{\gamma}_{12} M_z \left(P_x \nabla_z M_y - P_y \nabla_z M_x\right) +$$

$$+ \tilde{\gamma}_{13} \left(P_x \nabla_y - P_y \nabla_x\right) M_z^2 + \tilde{\gamma}_{14} \left(P_x M_y - P_y M_x\right) \nabla_z M_z \qquad (4.5)$$

For the case of dihedral groups ($mmm, mm2, 222$) of the orthorhombic crystal system, the free energy of inhomogeneous ME interactions is described by the expression:

$$F_{ME}^{D}=\Gamma_{1ii}P_i\,\nabla_i M_i^2+\Gamma_{2ij}P_i\,\nabla_i M_j^2+\Gamma_{3ij}P_i M_j\,\nabla_j M_i+\Gamma_{4ij}P_i M_i\,\nabla_j M_j \quad (4.6)$$

In order to simplify the expression, we renormalized the constants of the inhomogeneous ME interaction. Also, to simplify the writing, the final form of the expression is not represented as the traditional sum of symmetric (1.6) and antisymmetric (1.5) terms.

It is important to classify the symmetry groups of the paramagnetic phase as: symmetric ($S$), alternating ($A$), dihedral ($D$) and cyclic ($C$) groups [39]. These purely algebraic properties directly determine the invariant of inhomogeneous ME interactions: $F_{ME}^{S}$ (4 phenomenological constants), $F_{ME}^{A}$ (7 constants), $F_{ME}^{D}$ (11 and 21 constants in tetragonal and orthorhombic crystal systems, respectively) and $F_{ME}^{C}$ (21 constants), respectively. The invariant $F_{ME}^{D}$ of an orthorhombic crystal contains 4 tensors with interaction constants.

At the same time, centrosymmetric, polar, and enantiomorphic types of symmetry groups do not directly correlate with free energy expressions, as was the case for the Dzyaloshinskii-Moriya interaction. Accordingly, the chirality of a magnetically ordered medium does not determine the presence of an inhomogeneous ME interaction, which is consistent with the conclusion of [35], according to which such an interaction is possible in any magnetically ordered medium (including centrosymmetric, enantiomorphic, or polar).

The proposed phenomenological expressions can be easily generalized to the cases of antiferromagnetic ordering according to the scheme described for $Cr_2BeO_4$ (two antiferromagnetic vectors) [121] and multiferroics $BiFeO_3$ (one antiferromagnetic vector) [80].

Component (1.4) turns out to be fundamental in $YMn_2O_5$ [82]. In general, for ferromagnets, this part of the inhomogeneous ME interaction can be studied experimentally in the vicinity of the compensation point. Also, a sufficiently strong external electric field can create a polarization that will change the magnetization modulus due to the ME interaction. Such an approach can be used experimentally to establish and/or measure the interaction constants due to the component (1.4) in a given crystal. According to this phenomenological theory, the term $\tilde{\gamma}_2(P\,\nabla)M^2$ exists only in the $m\bar{3}m$ the crystal. An orthorhombic crystal $YMn_2O_5$ has invariants of the form $\Gamma_{2ij}P_i\nabla_i M_j^2$ which induce a similar mean electric polarization distribution. In general, the properties of inhomogeneous ME interactions $m\bar{3}m$ crystal are preserved in other cubic, tetragonal, and orthorhombic crystals. The crystallographic groups of these crystals are subgroups of $m\bar{3}m$. Accordingly, additional components of the induced polarization arise at similar ME effects. Thus, in order to group the free energy invariants responsible for similar ME effects, the numbering of the phenomenological constants was retained here for different crystal systems.

At the same time, new ME properties arise in crystals of cyclic groups ( $4_z/m_z$, $\bar{4}_z$, $4_z$) of tetragonal symmetry, since additional types of phenomenological invariants arise, including Lifshitz invariants. Accordingly, inhomogeneous ME interactions in crystals of cyclic crystallographic groups deserve attention in experimental studies, since new types of ME effects can arise (new components of electric polarization), which are described by additional 10 constants ($\tilde{\gamma}_5$ - $\tilde{\gamma}_{14}$) in tetragonal crystals.

Thus, the invariants $(1-\delta_{iz}-\delta_{jz})\varepsilon_{ijk}\tilde{\gamma}_8 P_j\nabla_i M_i^2$ and $\tilde{\gamma}_{14}(P_x M_y - P_y M_x)\nabla_z M_z$ invariants induce similar phenomena in a similar experimental setup [31], where the appearance of a normal (to the film) polarization component was observed. The new phenomena in a tetragonal crystal will be the induction of electric polarization in the film plane (homogeneous in the DW plane), which leads to the deformation of the Néel DW under the action of the applied electric potential of the electrode in the form of a point. Due to the presence of polarization in the plane, complex bends of the DW arise, in contrast to the simple shear that has already been observed. For a qualitative description and prediction of these ME effects in micromagnetic structures, the symmetry classification presented here is sufficient.

A specific example of a material in which these ($\tilde{\gamma}_8$ and $\tilde{\gamma}_{14}$) ME invariants play a role are crystals of tetragonal cyclic crystallographic groups, e.g. $FeNb$ with a space group:

$$G = I4/m1' \tag{4.7}$$

in the paramagnetic phase. An indirect approach to experimental verification is the electro-magneto-optical effect (i.e., Faraday rotation induced by an electric field) in the vicinity of magnetic DWs, which was observed in Yttrium iron garnet films [30].

The point groups of hexagonal and trigonal crystals are cyclic or dihedral. Accordingly, the structure of inhomogeneous ME interactions is expected to be similar for uniaxial crystals. The corresponding expression for the free energy always has a peculiarity for the $z$-component (4.4-4.5). Thus, the above-mentioned ME effects based on magnetic DWs (described by the constants $\tilde{\gamma}_8$ and $\tilde{\gamma}_{14}$) are expected in optically uniaxial crystals. On the contrary, these effects will not occur in optically isotropic crystals.

Chiral structures can be induced by inhomogeneous ME interaction in any magnetically ordered medium. The collapse of a magnetic skyrmion to the atomic scale requires a microscopic analysis (similar to [91-92]) of the ME interaction. The skyrmions can be induced by any Dzyaloshinskii-Moriya interaction: intrinsic chiral medium interaction or inhomogeneous ME interaction. In the first case, the Dzyaloshinskii-Moriya vector is determined by the crystallographic axes. In the second case, it depends on the vectors $\boldsymbol{e}_{12}$ and $\boldsymbol{P}$ according to the symmetry requirements. Since the magnetic moment is an axial t-odd vector, the vector product $[\boldsymbol{S}_1 \times \boldsymbol{S}_2]$ is an axial t-even vector. The vector product $[\boldsymbol{e}_{12} \times [\boldsymbol{S}_1 \times \boldsymbol{S}_2]]$ is, respectively, a t-even polar vector. Finally, the invariant of the free energy density of a crystal is:

$$I_0 = \boldsymbol{P}[\boldsymbol{e}_{12} \times [\boldsymbol{S}_1 \times \boldsymbol{S}_2]], \tag{4.8}$$

that can exist in a crystal of any crystallographic symmetry.

The polarization vector (1.7) is obtained from this invariant according to (4.1). Nevertheless, the triad $M-O-M$ is a part of the crystal lattice, so its physical properties must satisfy the symmetry requirements of the spatial symmetry group, which determines the structure of the microscopic interaction tensor $\hat{d}$. After the transition to macroscopic quantities and continuous spatial analysis of the crystal, these microscopic ME interactions are averaged and the final expression (1.5) is determined by the point group of magnetic symmetry of the crystal $G_P$. It should be noted that this is exactly the inhomogeneous [122] part of the Dzyaloshinskii-Moriya interaction. The anisotropic invariants (including antisymmetric ones) retain their form after the transition to the macroscopic level of description. Thus, the next largest component of expression (1.8) is the anisotropic symmetric interaction (indirect exchange of the pseudodipole type) $S_1 \hat{\Gamma} S_2$. Its relativistic type means that it is quadratic with respect to the spin-orbit interaction in the expansion for the indirect exchange interaction.

## 4.2. Inhomogeneous magnetoelectric effects in magnetic domain walls

Local inhomogeneous ME effects in magnetic DWs of a cubic $m\bar{3}m$ crystal are usually described by the following expression [33]:

$$\boldsymbol{P}=\gamma_0\left[\left(\boldsymbol{M}\,\nabla\right)\boldsymbol{M}-\boldsymbol{M}\left(\nabla\boldsymbol{M}\right)+\ldots\right] \tag{4.9}$$

In the case of a planar DW, expression (4.9) takes a one-dimensional form:

$$\boldsymbol{P}=\gamma_0\left(M_z\frac{\partial\boldsymbol{M}}{\partial z}-\boldsymbol{M}\frac{\partial M_z}{\partial z}\right) \tag{4.10}$$

Thus, only the transverse polarization components are non-zero:

$$P_x=\gamma_0\left(M_z\frac{\partial M_x}{\partial z}-M_x\frac{\partial M_z}{\partial z}\right) \tag{4.11}$$

$$P_y=\gamma_0\left(M_z\frac{\partial M_y}{\partial z}-M_y\frac{\partial M_z}{\partial z}\right) \tag{4.12}$$

It is necessary to analyze the relationship between expression (4.10) and the group-theoretical predictions. If the function $M_i(z)$ is odd, then $\frac{\partial M_i(z)}{\partial z}$ is even, where $i=x,y,z$. If the function $M_i(z)$ is even, then $\frac{\partial M_i(z)}{\partial z}$ is odd. The rules for products of even and odd functions are simple. These rules allow us to qualitatively describe the spatial distribution of the electric polarization induced by the inhomogeneous ME interaction. The differences between these results and the group-theoretical predictions are that the component $P_z(z)$ component is always zero according to (4.10), but is a permitted symmetry (Tables 2.1-2.4). The components $P_x(z)$ and $P_y(z)$ of selected achiral DWs with one magnetization component ($|\boldsymbol{M}| \neq const$) have the same properties: $k=3,4,6,44$ and 45 (Table 2.4). All other point groups have a complete coincidence between the symmetric predictions and the expression (4.10) for the components $P_x(z)$ and $P_y(z)$.

Zero (according to (4.9)) components of the electric polarization (allowed by symmetry) can be induced by interactions of a form different from (4.9): other ME interaction invariants (equations (1.4), (1.6) and (4.2-4.6)), which are allowed by symmetry $m\bar{3}m$, nonlocal ME interaction (e.g., ME coupling due to mechanical stresses), the influence of free surfaces of the sample on the ME properties, etc.

Different couples of groups $G_k$ and $G_B$ ($G_k \subseteq G_B$) correspond to different types of degeneration and/or different degrees of degeneracy $q_k$ DW. There are two types of structural transitions: with and without preservation of the point group of magnetic symmetry. Transitions between DW states with the same symmetry $G_k$ and energy can be caused by an external influence that is not an invariant of all symmetry operations $g_i^{(l)}$ that correlate these two DWs. Changes of DW symmetry correspond to all possible subgroups $G_k$ groups. $G_B$:

$$1 \subseteq G_k \subseteq G_B \tag{4.13}$$

Thus, the study of relations of the type (4.13) for the entire list of groups $1 \leq k \leq 64$ (Tables 2.1-2.4) leads to the possibility of establishing possible structural transitions (for the magnetic and electronic subsystems) in arbitrary magnetic

DWs. DWs with different symmetry groups have different energies. Accordingly, transitions with a change of DW symmetry can be a relaxation from a metastable state to a state with a lower energy. An example is the transformation of the Néel DW into the Bloch limit after switching off the external magnetic field in the film plane [31].

The existing experimental observations of the inhomogeneous ME effect in DW [31] are consistent with the presented symmetry theory. The corresponding experimental sample has a crystallographic orientation (210) and contains equilibrium DWs with the plane oriented parallel to (001). Accordingly, the point group of magnetic symmetry of the sample surface $G_S$ has the form:

$$G_S = \infty_{[210]}/mmm1' \quad (4.14)$$

After its intersection with $G_P^{\infty} = m\bar{3}m1'$ one gets:

$$G_P = 2_{[001]}/m1' \quad (4.15)$$

The point group of boundary conditions should be a subgroup of the latter. The directions of magnetization of the domains are collinear to the crystallographic directions of type <210>. Therefore, the corresponding group has the form ($k=5$, Table 2.4):

$$G_B = 2_z'/m_z \quad (4.16)$$

Subgroups $G_k$ of this group describe the symmetry of possible DWs that can exist under fixed boundary conditions (domains). The corresponding groups are: $2_z'$, $\bar{1}'$ and $m_z$. The case of $G_k = 2_z'$ case describes a Bloch DW with an antisymmetric magnetization distribution. The DW with $G_k = \bar{1}'$ exists only in the vicinity of phase transitions. The DW with $G_k = m_z$ ($k=11$, table 2.3) has an even dependence $M_z(z)$ and odd dependencies of the transverse components $M_x(z)$ and $M_y(z)$. Such symmetry requirements correspond to the DW observed in the study [31]. The functions $P_x(z)$ and $P_y(z)$ functions in the volume of this DW are even, which corresponds to the uncompensated charges on the free surfaces of the sample in the above-mentioned experiment. The corresponding dipole properties were found in a strong electric field gradient. These DWs have simultaneous ferromagnetic and ferroelectric degeneration. Since these DWs have only 2 states (degree of degeneracy $q_W^{'}=2$), this combined degeneration allows ME properties. The change of state corresponds to a simultaneous change of the magnetic and electronic subsystems. At the same time, a DW shift is recorded (the effect sign [31]) and the corresponding states are connected by lost operations $g_i^{(l)}=2_z^{'}$ or $\bar{1}'$ in this way:

$$G_B=G_k+\bar{1}^{'}G_k=G_k+2_z^{'}G_k \tag{4.17}$$

The magnetic field in the sample plane and the electric field gradient are not invariants of these lost operations, so they switch DW states. The hypothetical (allowed by symmetry) odd component $P_z(z)$ in this DW means the existence of the bound charge distribution already in the DW volume. Obviously, this can be established by applying a strong homogeneous electric field, which has not yet been done experimentally. This type of experiments will allow one to test the interaction of the type (1.4) and (1.6).

Also, the inhomogeneous ME interaction of the type (1.4) can be detected by means of a strong electric field gradient applied to an achiral magnetic DW with a collinear magnetization distribution: $k=3,4,6,44$ and 45 (Table 2.4).

The above analysis of inhomogeneous ME interactions in the volume of magnetic DWs can be easily generalized to the cases of weak ferromagnets with non-collinear magnetization distribution and antiferromagnets using the algorithms proposed in [123] and [37], respectively. Also, the analysis is easily adapted to the case of moving DWs taking into account symmetry properties of the velocity vector (polar t-odd one) [37].

## 4.3. Types of variational problem and free energy of inhomogeneous magnetoelectric effects

In general, according to Neumann's principle, the symmetry of physical properties can be higher than the symmetry of the medium. Nevertheless, in the presence of different (including unknown yet) physical mechanisms of interaction in a solid state crystal, it is most likely that phenomenological variational calculations and qualitative symmetry predictions will coincide exactly. The differences are explained by idealizations and experimental setup-related measurement precision. In the present case, it is also necessary to check the completeness of the model in the sense of its idealizations or even the logical sequence of model construction.

For the case of arbitrary (with any possible point magnetic symmetry) planar magnetic DWs, one was compared the results of the symmetry analysis and the study of the equilibrium spatial distribution of the electric polarization obtained from expression (4.9) with one phenomenological constant of the ME interaction, which is most common in the bibliography. The results of the comparison show that the coincidence of these two alternative methods of analysis occurs for most transverse components of electric polarization. For the normal component $P_z$ the situation is the opposite. Magnetic symmetry allows the existence of this component for all 64 point groups of magnetic symmetry of magnetic DWs. At the same time, expression (4.9) always forbids the normal component. Thus, the fundamental symmetry theory [9] contradicts the expression (4.9), which is familiar to the modern bibliography. In this section, one sets out to identify the reasons for such contradictions.

The expression (4.9) of the Lifshitz invariant type was obtained by removing the full spatial derivatives of the form $\nabla_k M_j M_n$ from the free energy density, since they do not affect the solution of the variational problem. This approach is completely consistent for variational problems of the unknown function type $\boldsymbol{M}(r)$. At the same time, ME interactions have not been considered. One can call this type of variational problem as an *M-problem,* in which $\boldsymbol{P}(r)$ is considered as a fixed (known) parameter. An example of a real problem of this type is the helicoidal magnetic structure of orthorhombic antiferromagnets induced by an electric field. Also, the problem of electrostatic switching of a magnetic vortex into an antivortex [124] and the control of the micromagnetic structure of magnetic DWs by an electric field [31] were considered within the framework of the M-problem. The use of this approach yielded positive results compared to the experiment where changes in the micromagnetic structure (magnetic DW shift) were considered under the influence of and depending on an external electric field. The fixed parameter $\boldsymbol{P}(r)$ for cases of this type means that the electric polarization is determined *only by the* external electric field:

$$\boldsymbol{P}_0(r)=\chi_e \boldsymbol{E}(r), \tag{4.18}$$

where here and hereinafter the scalar dielectric constant is considered for simplicity.

Thus, an additional requirement for the consistency of the M-problem in general is that the electric polarization $\boldsymbol{P}_{ME}(r)$ induced by the inhomogeneous ME interaction must be significantly less than $\boldsymbol{P}_0(r)$ (comparing magnitudes).

In the case when the unknown function of the variational problem is the dependence $\boldsymbol{P}(r)$ (*P-problem*) or both $\boldsymbol{M}(r)$ and $\boldsymbol{P}(r)$ (*MP-problem*), the approach of removing the terms $\nabla_k M_j M_n$ from the free energy density of inhomogeneous ME interactions turns out to be erroneous, since the Euler-Lagrange equations with respect to the unknown function $\boldsymbol{P}_{ME}(r)$ depends on these terms, which affects the solution of the entire variational problem. Thus, the known expression for the energy of inhomogeneous ME interactions in a cubic $m\bar{3}m$ crystal with a single phenomenological constant of inhomogeneous ME interactions is inaccurate in the general case.

The following invariants remain constant after all the symmetry operations of the point group of magnetic symmetry $m\bar{3}m1'$ applied:

$$I_1=P_i M_i \nabla_i M_i, \tag{4.19}$$

$$I_2=P_i M_j \nabla_i M_j (1-\delta_{ij}), \tag{4.20}$$

$$I_3=P_i M_j \nabla_j M_i (1-\delta_{ij}), \tag{4.21}$$

$$I_4=P_i M_i \nabla_j M_j (1-\delta_{ij}), \tag{4.22}$$

The symmetry transformations of the group $m\bar{3}m1'$ do not transform these invariants into each other. Thus, they must be present in the expression for the free energy with different phenomenological constants. The M-problem corresponds to an unknown function $\boldsymbol{M}(r)$ and a fixed electric polarization. In this case, the invariants (4.19-4.20) are the full spatial derivatives of the unknown functions. In the case of the M-problem, these terms can be removed from the free energy, since they turn to zero in the Euler-Lagrange equations. After removing the full derivatives (the symmetric part) from the sum of equations (4.19-4.22) with constants, they are transformed into an expression with one phenomenological constant. For the M-problem, the free energy of inhomogeneous ME interactions takes the known form:

$$F_{ME}=\gamma_0 \boldsymbol{P}\left[\boldsymbol{M}(\nabla \boldsymbol{M})-(\boldsymbol{M}\,\nabla)\boldsymbol{M}\right] \tag{4.23}$$

A problem of type MP corresponds to the most general problem. The full derivatives $\boldsymbol{M}(r)$ affect the solution of the variational problem and cannot be removed from the Euler-Lagrange equations. Expression (4.9) can be considered only as an approximate description of the phenomena for such problems.

The general form of inhomogeneous ME interactions in a cubic $m\bar{3}m$ crystal, which can be used in all problems (M-, P-, and MP-problems), is given by the following expression with four phenomenological constants:

$$F_{ME}=\gamma_1 I_1+\gamma_2 I_2+\gamma_3 I_3+\gamma_4 I_4 \tag{4.24}$$

Based on (2), we obtain:

$$F_{ME}=\gamma_1 I_1+\gamma_2\left[\frac{(\boldsymbol{P}\,\nabla)\boldsymbol{M}^2}{2}-I_1\right]+\gamma_3\left[\boldsymbol{P}(\boldsymbol{M}\,\nabla)\boldsymbol{M}-I_1\right]+\gamma_4\left[(\boldsymbol{P}\boldsymbol{M})(\nabla \boldsymbol{M})-I_1\right] \tag{4.25}$$

It is convenient to introduce new phenomenological constants:

$$\tilde{\gamma}_1=\gamma_1/2-\gamma_2-\gamma_3-\gamma_4 \tag{4.26}$$

$$\tilde{\gamma}_2=\gamma_2/2 \tag{4.27}$$

$$\tilde{\gamma}_3=\gamma_3 \tag{4.28}$$

$$\tilde{\gamma}_4 = \gamma_4 \tag{4.29}$$

Using these constants, the final expression of the energy of inhomogeneous ME interactions is as follows:

$$F_{ME} = \tilde{\gamma}_1 P_i \nabla_i M_i^2 + \tilde{\gamma}_2 (\boldsymbol{P}\nabla) \boldsymbol{M}^2 + \tilde{\gamma}_3 \boldsymbol{P}(\boldsymbol{M}\nabla)\boldsymbol{M} + \tilde{\gamma}_4 (\boldsymbol{P}\boldsymbol{M})(\nabla \boldsymbol{M}) \tag{4.30}$$

The second term in (4.30) describes the type of inhomogeneous ME interactions that exist only with a spatially modulated magnetization modulus [19]. This type of interaction is possible in the vicinity of the Curie temperature, as well as in the case of an elementary magnetic cell of a crystal containing more than one type of magnetic ions.

The fourth term (4.30) exists only in micromagnetic structures with demagnetization fields. It is necessary to compare expressions (4.9) and (4.30). The material constants of these expressions are related:

$$\gamma_0 = \frac{\tilde{\gamma}_3 - \tilde{\gamma}_4}{2} \tag{4.31}$$

The type of inhomogeneous ME interactions described by the constants $\tilde{\gamma}_1$ and $\tilde{\gamma}_2$ do not affect the equilibrium electronic and magnetic structures of the crystal in the case of the M-problem.

The free energy density (4.30) determines the direction of electric polarization $\boldsymbol{P}_{ME}$ which is induced by the material in the absence of external influences. The modulus and direction of this polarization is determined in the same way as in the case of a dielectric material in an external electric field:

$$\boldsymbol{P}_{ME} = \chi_e \boldsymbol{E}_{ME}, \tag{4.32}$$

The effective field of inhomogeneous ME interactions is given by the following expression:

$$\boldsymbol{E}_{ME} = -\partial F_{ME} / \partial \boldsymbol{P} \tag{4.33}$$

The general polarization is as follows:

$$\boldsymbol{P}=\chi_e\left[\boldsymbol{E}+\tilde{\gamma}_1 e_i \nabla_i\left(\boldsymbol{M}_i^2\right)+\tilde{\gamma}_2 \nabla\left(\boldsymbol{M}^2\right)+\tilde{\gamma}_3\left(\boldsymbol{M}\,\nabla\right)\boldsymbol{M}+\tilde{\gamma}_4\boldsymbol{M}\left(\nabla\boldsymbol{M}\right)\right] \tag{4.34}$$

Let us consider the P-type variational problem. Within the one-dimensional model of a plane magnetic DW without the presence of an external electric field, expression (4.34) takes the form:

$$\boldsymbol{P}=\chi_e\left[\tilde{\gamma}_1\boldsymbol{e}_z\frac{\partial M_z^2}{\partial z}+\tilde{\gamma}_2\boldsymbol{e}_z\frac{\partial \boldsymbol{M}^2}{\partial z}+\tilde{\gamma}_3 M_z\frac{\partial \boldsymbol{M}}{\partial z}+\tilde{\gamma}_4\boldsymbol{M}\frac{\partial M_z}{\partial z}\right], \tag{4.35}$$

where the term $\tilde{\gamma}_2\partial\boldsymbol{M}^2/\partial z$ can be removed if the crystal is not in a state close to the Curie point. The following expressions give the polarization components:

$$P_x=\chi_e\left[\tilde{\gamma}_3 M_z\frac{\partial M_x}{\partial z}+\tilde{\gamma}_4 M_x\frac{\partial M_z}{\partial z}\right] \tag{4.36}$$

$$P_y=\chi_e\left[\tilde{\gamma}_3 M_z\frac{\partial M_y}{\partial z}+\tilde{\gamma}_4 M_y\frac{\partial M_z}{\partial z}\right] \tag{4.37}$$

$$P_z=\chi_e\left[\left(\tilde{\gamma}_1+\frac{\tilde{\gamma}_3+\tilde{\gamma}_4}{2}\right)\frac{\partial M_z^2}{\partial z}+\tilde{\gamma}_2\frac{\partial M^2}{\partial z}\right] \tag{4.38}$$

Expression (4.38) shows that the function $P_z(z)$ is odd when the functions $\boldsymbol{M}(z)$ or $M_z(z)$ are both even or odd. All these functions can be of type (*F*) only simultaneously. The point group of magnetic symmetry of a DW cannot correspond to a symmetry higher than the crystal symmetry. In general, this group depends on the orientation of the DW plane relative to the crystal surface. The effect of free sample surfaces on the DW symmetry is automatically described by the group-theoretical analysis (Section 3). This automatically takes into account the effect of the electric field of the sample surface bound charges on the DW structure.

All DWs are achiral if they have this property (this is not limited to achiral DWs):

$$M_z=0 \tag{4.39}$$

$$\partial\boldsymbol{M}^2/\partial z\neq 0 \tag{4.40}$$

Domain walls of the Bloch type have a small or zero (with strict equality $\partial \boldsymbol{M}^2/\partial z = 0$) longitudinal electric polarization. They include DWs with a center:

$$G_k = 2_x' 2_y 2_z', \tag{4.41}$$

Bloch DWs without a center:

$$G_k = 2_z' \tag{4.42}$$

and Bloch 0° DW with symmetry group:

$$G_k = 2_z'/m_z' \tag{4.43}$$

Néel domain walls always have a longitudinal electric polarization with a characteristic value:

$$P_z \approx \chi_e M_S^2 \left[\tilde{\gamma}_1 + \left(\tilde{\gamma}_3 + \tilde{\gamma}_4\right)/2\right]/\delta_0, \tag{4.44}$$

where $\delta_0$ is the thickness of the DW. These include Néel DWs without a center:

$$G_k = m_x' \tag{4.45}$$

"charged" DWs of Néel with the center:

$$G_k = m_x' m_z' 2_y \tag{4.46}$$

and the Néel 0°-DW with the symmetry group:

$$G_k = 2_x'/m_x', \tag{4.47}$$

as well as uncharged Néel DWs with a center:

$$G_k = m_z m_y' 2_x', \tag{4.48}$$

Thus, the new phenomenological approach yields virtually complete agreement with the symmetry predictions. Among the point groups of magnetic symmetry of DW, 47 groups out of 52 (a subset of 64 groups for a cubic crystal) show a complete coincidence of the magnetization and electric polarization distributions predicted by symmetry and calculated from the phenomenological theory. The remaining 5 groups describe the DW symmetry with one magnetization component and two electric polarization components. A complete coincidence of predictions is possible if other types of inhomogeneous ME interactions are taken into account.

A common feature of these phenomenological results (based on the four-constant approach) is the existence of an odd component of the longitudinal electric polarization. This is a difference from the one-constant phenomenology. This corresponds to the distribution of the bound electric charge in the DW volume, which can be detected experimentally in the form of the sensitivity of the micromagnetic structure to an external homogeneous electric field. Such an experiment will be free from the influence of the electric dipole in the magnetic DW, which has already been established experimentally, which should be considered as a verification of the four-constant phenomenology.

### 4.4. Inhomogeneous magnetoelectric effects in Bloch lines

The presented symmetry-based BL systematization (Tables 2.5-2.8) describes ME effects in crystals of any symmetry. The corresponding phenomenological theory is proposed above (Section 4.1). These two results need to be compared, for which we will consider a crystal of the highest symmetry, which will guarantee symmetric predictions in crystals with crystallographic groups that are subgroups of $m\bar{3}m$. Indeed, higher symmetry prohibits certain invariants of the inhomogeneous ME interaction. The transition to subgroups (lower symmetry) leads to the removal of these restrictions, which corresponds to the symmetric predictions of the BL structure, which contain fundamentally possible components of the order parameters in the BL volume.

The electric polarization induced by the inhomogeneous ME effect in a cubic $m\bar{3}m$ crystal is described by the following 4-constant expression:

$$\boldsymbol{P}=\chi_e\left[\tilde{\gamma}_1 \boldsymbol{e}_i \nabla_i\left(M_i^2\right)+\tilde{\gamma}_2 \nabla\left(\boldsymbol{M}^2\right)+\tilde{\gamma}_3(\boldsymbol{M}\nabla)\boldsymbol{M}+\tilde{\gamma}_4 \boldsymbol{M}(\nabla\boldsymbol{M})\right] \quad (4.49)$$

For ordinary micromagnetic variational problems (constant magnetization modulus), the second term is zero. Within the framework of the problem of BL in the volume of a planar magnetic DW, expression (4.49) takes the form:

$$P_x=\chi_e\left[\left(\tilde{\gamma}_1+\frac{\tilde{\gamma}_3+\tilde{\gamma}_4}{2}\right)\frac{\partial M_x^2}{\partial x}+\tilde{\gamma}_3 M_z\frac{\partial M_x}{\partial z}+\tilde{\gamma}_4 M_x\frac{\partial M_z}{\partial z}\right] \tag{4.50}$$

$$P_y=\chi_e\left[\tilde{\gamma}_3\left(M_x\frac{\partial M_y}{\partial x}+M_z\frac{\partial M_y}{\partial z}\right)+\tilde{\gamma}_4 M_y\left(\frac{\partial M_x}{\partial x}+\frac{\partial M_z}{\partial z}\right)\right] \tag{4.51}$$

$$P_z=\chi_e\left[\left(\tilde{\gamma}_1+\frac{\tilde{\gamma}_3+\tilde{\gamma}_4}{2}\right)\frac{\partial M_z^2}{\partial z}+\tilde{\gamma}_3 M_x\frac{\partial M_z}{\partial x}+\tilde{\gamma}_4 M_z\frac{\partial M_x}{\partial x}\right] \tag{4.52}$$

A simple analysis of the parity and oddness of the components shows that expressions (4.50-4.52) correspond to the symmetric predictions (Tables 2.5-2.8). This statement is correct for 40 point groups of BLs, which correspond to at least two non-zero magnetization components. The remaining 8 groups describe the BL symmetry with one magnetization component (possible only in the 0° DW volume) and two electric polarization components, which can occur near phase transitions. A complete correspondence between the phenomenological theory and the symmetry analysis arises if one takes into account an additional type of inhomogeneous ME interactions, which (the corresponding invariant) depends on the spatial derivatives of the electric polarization vector. From the point of view of the manifestation of such an interaction, the study of the ME properties of BLs in the vicinity of phase transitions is of fundamental importance.

A similar experimental setup [37], which was developed to detect the displacement of DW, can be used for studies of the displacement of BL. The corresponding BL should be with point groups:

$$U_l=2_x^{'}2_y2_z^{'}\left(2_y^{'}2_x2_z^{'}\right), \tag{4.53}$$

corresponding to the Bloch DW volume, with the component $M_z(z,x)\neq 0$ component exists only in the BL volume. The magnetic DW shifts under the influence of the electric field gradient were observed only for the Néel DW, while the Bloch domain walls do not generate transverse electric polarization components. Therefore, the scheme proposed here for checking the BL shift appears to be free from extraneous influences of the shift of the DW itself with BL, which ensures higher purity of the experiment. Preliminary experimental evidence of this type has already been published [32].

All lost symmetry operations belong to the point group of magnetic symmetry of the paramagnetic phase:

$$u_i^{(l)} \subset G_P \tag{4.54}$$

The action of these operations on an ordered medium does not change its total free energy. If there is some external influence, it can be changed by these operations as well. Considering the sources of this influence (for example, external charges or currents), one can consider the entire system: the solid and the sources of external influences. The effect of any magnetic symmetry operation on such a system does not change the total energy of the system. This is the result of the isotropy of space and invariance with respect to time reversal for all physical interactions. The criterion of time reversal is fulfilled in thermodynamically equilibrium states, which corresponds to our analysis. If one defines a dissipative process as a "state", then the presented symmetric properties can already be violated. Thus, the total energy is not changed by the operations $u_i^{(l)}$ operations for equilibrium states. At the same time, if the external influence is an invariant of at least one operation $u_i^{(l)}$then the degenerate state of the BL cannot be switched by this external influence. If the external influence is not an invariant of all lost operations $u_i^{(l)}$ in the decomposition (2.16), then this influence removes the degeneracy of states. Such an influence (for example, homogeneous or inhomogeneous electromagnetic fields) can switch $Q_l'$ states of the BL under given boundary conditions. At the same time, the DW fragments remain fixed. Thus, this approach allows one to establish the magnetic memory technology concept based on BLs. If the design of the corresponding memory elements allows changing the structure of the DW itself, the total number of states of the BL increases to a value:

$$N_{BL} = Q_B Q_l^{'} \quad (4.55)$$

This increases the possible magnetic memory capacity.

In general, the analysis of group-subgroup relations for each micromagnetic structure is a fairly general tool. If the degeneration appears to be ferromagnetic, ferroelectric, and/or ferroelastic at the same time, then different methods of controlling memory cells or other spintronic devices are possible (due to ME or elastic effects).

0° DWs are soliton-type states. The symmetry classification for such DWs is proposed in Section 2.1. The symmetry classification presented in this paper also includes such BLs that split the same DW regions. Such BLs in this symmetry classification have the following property (their criterion):

$$U_B = G_k \quad (4.56)$$

Such soliton-type BLs can be created in the DW volume using the point application of a strong electric field gradient. After the field is turned off, such micromagnetic structures can stabilize as metastable structures, for example, due to the Dzyaloshinskii-Moriya interaction in a chiral medium. In this sense, establishing chirality for all BLs appears to be important for experimental research.

The presented analysis of the relation between the symmetry classification and inhomogeneous ME properties for BL turns out to be similar to the relation in the case of DW. The connection consists of the same results of symmetric predictions and conclusions of the phenomenological theory of inhomogeneous ME properties. The same properties remain for BPs. The experimental methods described above for detecting inhomogeneous ME effects in DWs, BLs, and BPs are similar and can be achieved by observing the displacement of these micromagnetic structures under the impact of strong homogeneous and inhomogeneous electric fields.

## 4.5. Inhomogeneous magnetoelectric effects in structures with vortex magnetization distribution

The structure of MV and MAV can be switched regardless of the sample shape by means of inhomogeneous ME effects, as shown in [124]. The electrode that can be used in this case is the tip of an atomic force microscope [124]. Skyrmions, antiskyrmions, MVs, and MAVs can be induced by inhomogeneous ME interactions. A relevant practical application exists for the case of 0° DWs, which allows the development of memory elements based on these DWs. Skyrmions and these DWs are soliton structures, so they can arise under similar conditions. ME control method is important for the development of modern electronic and spintronic devices.

The polarization induced by the inhomogeneous ME interaction in the case of $m\bar{3}m$ crystal in the most general form is as follows:

$$\boldsymbol{P}=\chi_e\left[\tilde{\gamma}_1\boldsymbol{e}_i\,\nabla_i\left(M_i^2\right)+\tilde{\gamma}_2\,\nabla\left(\boldsymbol{M}^2\right)+\tilde{\gamma}_3\left(\boldsymbol{M}\,\nabla\right)\boldsymbol{M}+\tilde{\gamma}_4\boldsymbol{M}\left(\nabla\boldsymbol{M}\right)\right] \tag{4.57}$$

Let us consider the vortex structure (2.29) in the plane of a thin film with the induced anisotropy of the easy-plane anisotropy type. The distribution of electric polarization in the sample plane is as follows:

$$\boldsymbol{P}_\perp=\chi_e\left\{\tilde{\gamma}_1\left[\boldsymbol{e}_x\,\nabla_x\left(M_x^2\right)+\boldsymbol{e}_y\,\nabla_y\left(M_y^2\right)\right]+\tilde{\gamma}_2\,\nabla_\perp\left(M^2\right)+\tilde{\gamma}_3\left(\boldsymbol{M}_\perp\,\nabla_\perp\right)\boldsymbol{M}_\perp+\tilde{\gamma}_4\boldsymbol{M}_\perp\left(\nabla_\perp\boldsymbol{M}_\perp\right)\right\}, \tag{4.58}$$

where "$\perp$" means components in the plane ($xy$). A schematic illustration of this distribution is shown for the typical cases $\tilde{\gamma}_3\tilde{\gamma}_4<0$ and $\tilde{\gamma}_1^{\,2}\ll\left|\tilde{\gamma}_3\tilde{\gamma}_4\right|$ [33] (Figure 4.1). Structures with $n=3$ and $n=-1$ have identical point groups for the component $\boldsymbol{P}_\perp$, i.e., the group of the form:

$$G_V=4'/m_z'm_\perp'\,m_\perp \tag{4.59}$$

They have similar magnetization distributions (Fig. 2.2 (a), (b)). At the same time, the electric polarization distributions are similar, but the bound charges in the center have opposite signs (Fig. 4.1 (a), (b)) [33].

Let us consider the normal magnetization component in the following form, which is chosen purely for qualitative symmetry analysis:

$$M_z = M_s e^{\frac{-\rho}{\rho_0}} \tag{4.60}$$

In the case of a constant magnetization modulus (saturation magnetization $M_s$), (4.60) leads to the following changes in (2.29):

$$\boldsymbol{M}_{\perp}(\rho,\varphi) = M_s \sqrt{1-e^{\frac{-2\rho}{\rho_0}}}\left(\boldsymbol{e}_x \cos(n\varphi + c\pi/2) + \boldsymbol{e}_y \sin(n\varphi + c\pi/2)\right), \tag{4.61}$$

The normal (to the film plane) polarization component is given by (at $|\boldsymbol{M}| = M_S$):

$$P_z = \chi_e \left\{ \tilde{\gamma}_3 (\boldsymbol{M}_{\perp} \nabla_{\perp}) M_z + \tilde{\gamma}_4 M_z (\nabla_{\perp} \boldsymbol{M}_{\perp}) \right\} \tag{4.62}$$

For ansatz (4.61), expression (4.62) takes the form:

$$P_z = \chi_e M_s^2 \frac{e^{\frac{-\rho}{\rho_0}} \cos[(n-1)\varphi + c\pi/2]}{\rho_0 \sqrt{1-e^{\frac{-2\rho}{\rho_0}}}} \left\{ \tilde{\gamma}_3 \left( e^{\frac{-2\rho}{\rho_0}} - 1 \right) + \tilde{\gamma}_4 \frac{\rho_0}{\rho} e^{\frac{-2\rho}{\rho_0}} \left[ \frac{\rho}{\rho_0} + \left( e^{\frac{2\rho}{\rho_0}} - 1 \right) n \right] \right\} \tag{4.63}$$

The criterion of $P_z = 0$ is $n = 1$ and $c = \pm 1$ corresponding to the symmetry group of the following form (Fig. 2.2 (c), Fig. 4.1 (a)):

$$G_C = \infty_z 2'_x 2'_{xy} \tag{4.64}$$

Accordingly, expression (4.63) satisfies the symmetric predictions:

$$2'_x \cdot P_z = -P_z \Rightarrow P_z = 0 \tag{4.65}$$

The total uncompensated charge on one surface of the crystal sample is as follows:

$$Q_{SF} = \int_0^{2\pi} d\varphi \int_0^{+\infty} \rho P_z d\rho \tag{4.66}$$

In contrast to the well-studied bulk charge in the bibliography $Q_V$ [33], the surface charge $Q_{SF}$ is localized in the structure's core. It is given by the following expression ($n = 1$):

$$Q_{SF} = \frac{\chi_e \rho_0 M_s^2 \pi^2 \cos\frac{c\pi}{2}}{4} \left(\tilde{\gamma}_4 - \tilde{\gamma}_3\right)\left(1 + \log 4\right), \tag{4.67}$$

If $n \neq 1$ or $c = \pm 1$, then $Q_{SF} = 0$. Thus, the surface bound charge $Q_{SF}$ vanishes in the case of structures with rotation order $n > 1$ and in the case of $n < 0$ (Fig. 4.2). The well-studied structure with $n = 1$ and $c = \pm 1$ cannot contain the total charge $Q_{SF}$. On the contrary, at $n = 1$ and $c = 0$ there is a non-zero charge $Q_{SF}$. However, such structures have large demagnetization fields and are usually unstable. Structures with $n = 1$ and $c \notin Z$ (Fig. 2.2 (b)) have smaller demagnetization fields and contain a charge $Q_{SF}$.

In the case of a thin film and a nanodot, these surface charges correspond to a dipole appearance:

$$\boldsymbol{p} = \boldsymbol{e}_z Q_{SF} h, \tag{4.68}$$

where $h$ is the characteristic thickness along $Z$. This dipole participates in the ME interaction with the magnetization component $M_z$. Switching the direction of $\boldsymbol{p}$ leads to a switching $M_z$ or chirality. Quantitative analysis of this relationship requires an exact solution of the variational problem. Structures with $n \neq 1$ cannot be sensitive to an inhomogeneous electric field.

As mentioned above, there are symmetry criteria that prohibit an electric dipole in some structures. To allow a dipole, the point symmetry group must be polar. The difference between a bulk bound charge $Q_V$ is that it cannot be prohibited by symmetry arguments. Indeed, the electric charge (point) symmetry represents the highest possible point magnetic symmetry:

$$G_{ch} = \infty\infty m 1' \tag{4.69}$$

According to Neumann's principle, this property is allowed in any medium, although the nature of the charges may be different. For example, quantum Hall skyrmions have both topological $n$ and electric charges $Q_V$ [125].

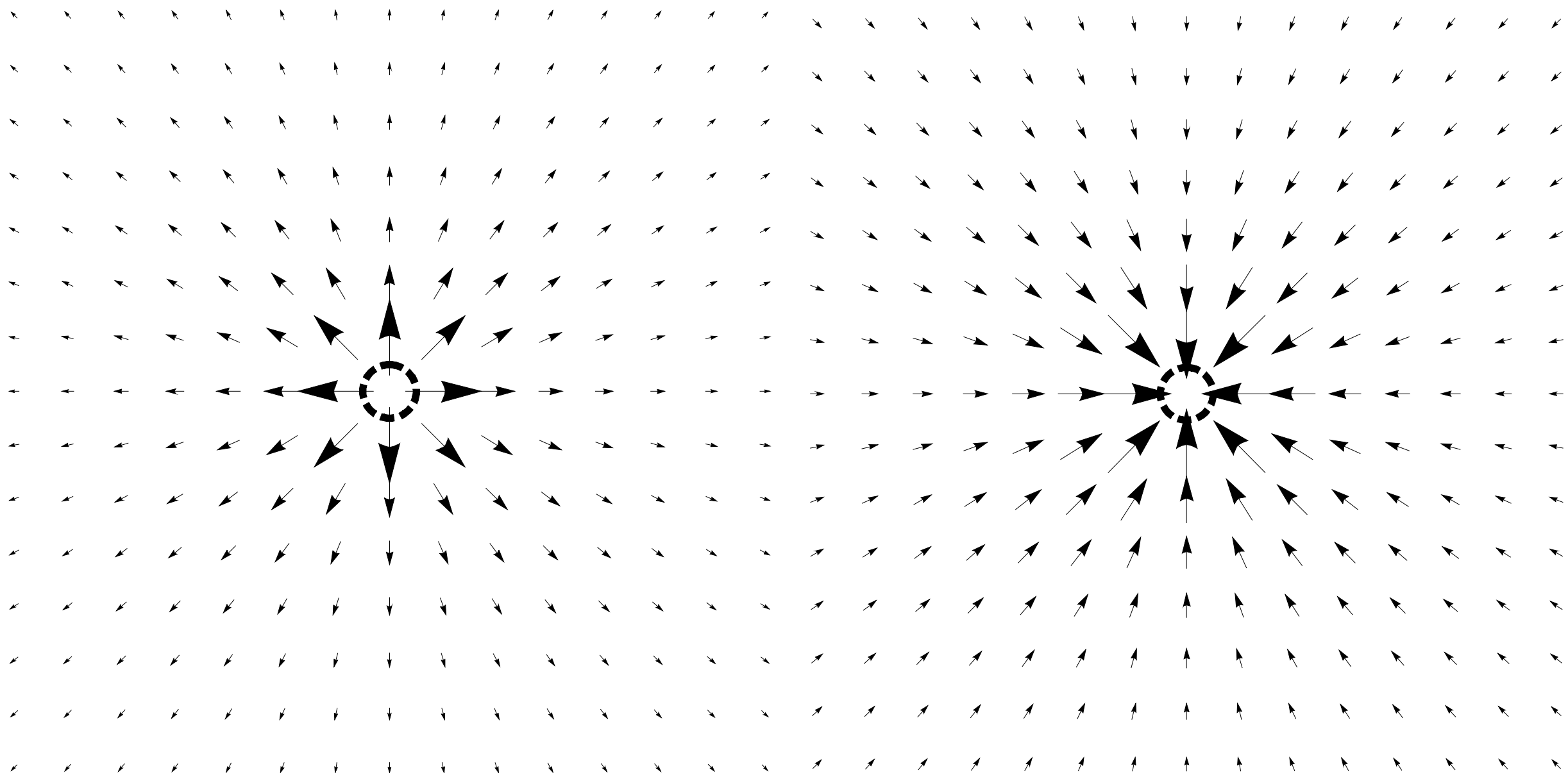

(a) (b)

**Fig. 4.1.** Schematic illustration of the electric polarization distribution in the sample plane for the cases $n>0$ (a) and $n<0$ (b).

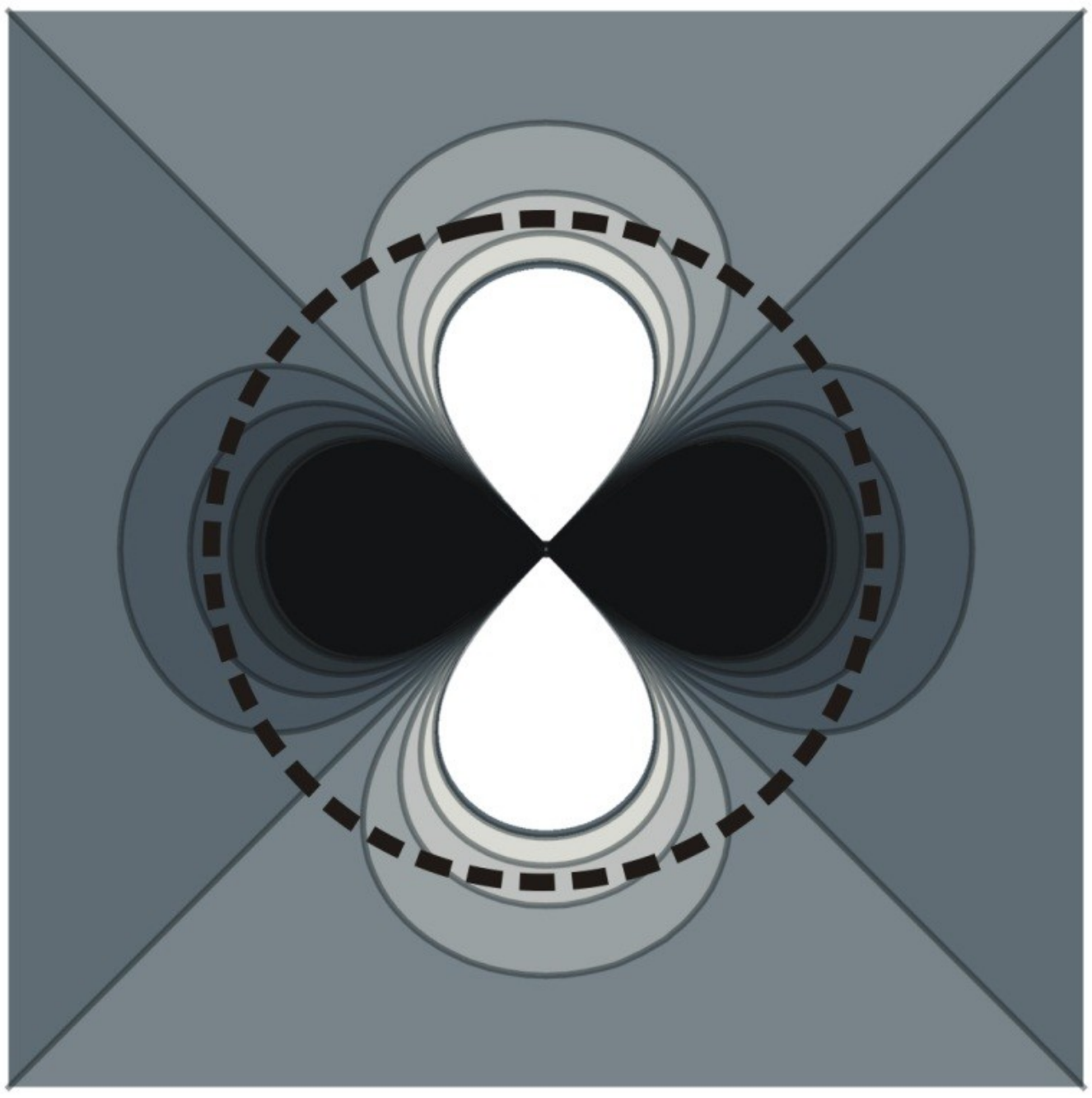

**Fig. 4.2.** Schematic illustration of the distribution of the normal component of the electric polarization for the case of the symmetry group $4'/m'_z m'_\perp m_\perp$ and vorticity $n=-1$. Black and white colors correspond to projections of opposite signs and equal modulus.

## 4.6. Conclusions

In this chapter, a phenomenological theory of inhomogeneous ME interaction in crystals of any symmetry of cubic, tetragonal, and orthorhombic crystal systems is proposed. The critical role of the type of crystallographic point group (symmetric, alternative, dihedral, or cyclic) for the nature of ME effects has been shown.

The phenomenological theory is compared with the results of the group-theoretic analysis for all typical micromagnetic structures with a solid positive result. The following conclusions have been made:

1. Tetragonal crystals with cyclic groups ($4_z/m_z$, $\bar{4}_z$ and $4_z$) have the most complex ME properties, which are described by the 21st phenomenological constants. Six different types of Lifshitz invariants are allowed by the symmetry of these crystals.

2. One of the possible tests of the proposed phenomenological theory is bending experiments with magnetic DWs. Some of the ME energy components can be tested in the vicinity of the ferromagnetism compensation point, where a sufficiently strong electric field of a point electrode can modulate the magnetization modulus.

3. A hypothetical type of inhomogeneous ME interaction (invariants of the form $M_i M_j \nabla_k P_l$), which provides a complete agreement between the symmetric predictions and the phenomenological theory, has been suggested.

4. The connection between the proposed phenomenological theory and various microscopic mechanisms of ME interactions, including the new (discovered in 2008) generalized Dzyaloshinskii-Moriya ME interaction, is established.

5. The central part of the vortex magnetization distribution contains an electric dipole induced by inhomogeneous ME interactions. It allows one to control the chirality of the corresponding structure as a state in possible memory devices.

6. Vortex structures with non-unitary vorticity do not contain an electric dipole.

7. All micromagnetic structures with t-odd chirality have qualitatively identical spatial distributions of magnetization and electric polarization.

8. New mechanisms of inhomogeneous ME interactions can be experimentally recorded by applying a strong homogeneous electric field (as opposed to the tip field that is usually used).

9. A new soliton-type BL can be created by the electric field of a point electrode on a magnetic DW.

# CONCLUSIONS

As a result of the theoretical studies carried out in this thesis, we obtained a complete symmetry classification of all the main micromagnetic structures, where the structures were first divided by their magnetic chirality. The organic connection of this classification with the inhomogeneous magnetoelectric effects in them has been shown. The inhomogeneous ME effect is qualitatively described for each symmetry class. The phenomenological theory of these effects is extended to the case of crystals of cubic, tetragonal, and orthorhombic symmetries. The obtained results are analyzed for their compatibility with the existing experimental facts and the possibility of detecting new phenomena. The following new results were obtained:

1. Micromagnetic structures with t-odd chirality have qualitatively identical spatial distributions $\boldsymbol{M}(r)$ and $\boldsymbol{P}(r)$.

2. In total, there are 64 groups of magnetic symmetry of planar magnetic DWs. Among them, 42 groups correspond to 180-degree DWs, 42 groups to 0-degree DWs, and 10 groups to other types of DWs.

3. In total, there are 48 groups of magnetic symmetry of BLs, including 22 (11 t-even and 11 t-odd) chiral and 26 achiral. The BP classification includes the same number of classes.

4. The connection between symmetry classification and degeneracy of micromagnetic structures is shown. Group-theoretical methods for studying the influence of the free surfaces of the sample on the spatial distributions of order parameters have been proposed.

5. The connection between the proposed phenomenological theory and various microscopic mechanisms of ME interactions, including the new (discovered in 2008) generalized Dzyaloshinskii-Moriya ME interaction, is established.

6. The important role of the type of crystallographic point group (symmetric, alternative, dihedral, or cyclic) in the nature of ME effects has been shown. These types uniquely determine the expression of the free energy of ME interactions.

7. The central part of the vortex magnetization distribution contains an electric dipole induced by inhomogeneous ME interactions. It allows one to control the chirality of the corresponding structure.

8. New mechanisms of inhomogeneous ME interactions can be experimentally recorded by applying a strong homogeneous electric field (as opposed to the tip field that is usually used).

9. BL and BP of the soliton type can be created by the electric field of a point electrode on a magnetic DW.

## LIST OF REFERENCES